\documentclass[reprint,superscriptaddress,amsmath,amssymb,aps,
 aps
floatfix,
]{revtex4-1}

\usepackage{graphicx}
\usepackage{dcolumn}
\usepackage{bm}
\usepackage{hyperref}
\usepackage{xcolor}
\usepackage{tikz}
\usepackage{xparse}
\usepackage{amsmath}
\usepackage[sc]{mathpazo} 
\usepackage[T1]{fontenc} 
\usepackage{microtype} 
\usepackage{amsmath}

\usepackage[font=small,labelfont=bf,justification=justified,format=plain]{caption} 
\usepackage{booktabs} 
\usepackage{float} 
\usepackage{hyperref} 
\restylefloat{figure}
\usepackage{graphicx}
\usepackage{array}
\newcolumntype{C}[1]{>{\centering\arraybackslash}p{#1}}

\usepackage{lettrine} 
\usepackage{paralist} 

\usepackage{fancyhdr} 
\usepackage{xcolor}
\usepackage{xparse,xcoffins}
\ExplSyntaxOn
\NewCoffin\imagecoffin
\NewCoffin\labelcoffin

\keys_define:nn { miguel/label }
 {
  label   .tl_set:N = \l_miguel_label_tl,
  labelbox .bool_set:N = \l_miguel_label_box_bool,
  labelbox .default:n = true,
  fontsize .tl_set:N = \l_miguel_label_size_tl,
  fontsize .initial:n = \footnotesize,
  pos .choice:,
  pos/nw .code:n = \tl_set:Nn \l_miguel_label_pos_tl { left,up },
  pos/ne .code:n = \tl_set:Nn \l_miguel_label_pos_tl { right,up },
  pos/sw .code:n = \tl_set:Nn \l_miguel_label_pos_tl { left,down },
  pos/ssw .code:n = \tl_set:Nn \l_miguel_label_pos_tl { left,downdown },
  pos/sww .code:n = \tl_set:Nn \l_miguel_label_pos_tl { leftleft,down },
  pos/se .code:n = \tl_set:Nn \l_miguel_label_pos_tl { right,down },
  pos/sse .code:n = \tl_set:Nn \l_miguel_label_pos_tl { rightright,downdown },
  pos/see .code:n = \tl_set:Nn \l_miguel_label_pos_tl { rightmed,down },
  pos/n .code:n = \tl_set:Nn \l_miguel_label_pos_tl { hc,up },
  pos/w .code:n = \tl_set:Nn \l_miguel_label_pos_tl { left,vc },
  pos/s .code:n = \tl_set:Nn \l_miguel_label_pos_tl { hc,down },
  pos/e .code:n = \tl_set:Nn \l_miguel_label_pos_tl { right,vc },
  pos .initial:n = nw,
  unknown .code:n   = \clist_put_right:Nx \l_miguel_label_clist
                       { \l_keys_key_tl = \exp_not:n { #1 } }
 }
\clist_new:N \l_miguel_label_clist
\box_new:N \l_miguel_label_box
\box_new:N \l_miguel_label_image_box

\NewDocumentCommand{\xincludegraphics}{O{}m}
 {
  \group_begin:
  \tl_clear:N \l_miguel_label_tl
  \clist_clear:N \l_miguel_label_clist
  \keys_set:nn { miguel/label } { #1 }
  \tl_if_empty:NTF \l_miguel_label_tl
   {
    \miguel_includegraphics:Vn \l_miguel_label_clist { #2 }
   }
   {
    \SetHorizontalCoffin\imagecoffin
     {
      \miguel_includegraphics:Vn \l_miguel_label_clist { #2 }
     }
    \SetHorizontalCoffin\labelcoffin
     {
      \raisebox{\depth}
       {
        \bool_if:NTF \l_miguel_label_box_bool
         { \fcolorbox{white}{white}{\l_miguel_label_size_tl\l_miguel_label_tl} }
         { \l_miguel_label_size_tl\l_miguel_label_tl }
       }
     }
    \SetVerticalPole\imagecoffin{left}{0.2\Width}
    \SetVerticalPole\imagecoffin{leftleft}{0.3\Width}
    \SetVerticalPole\imagecoffin{right}{0.8\Width}
    \SetVerticalPole\imagecoffin{rightmed}{0.9\Width}
       \SetVerticalPole\imagecoffin{rightright}{0.85\Width}
    \SetHorizontalPole\imagecoffin{up}{\Height-3pt-\CoffinHeight\labelcoffin/2}
    \SetHorizontalPole\imagecoffin{down}{0.25\Height}
    \SetHorizontalPole\imagecoffin{downdown}{0.2\Height}
    \use:x{\JoinCoffins\imagecoffin[\l_miguel_label_pos_tl]\labelcoffin[vc,hc]} 
    \TypesetCoffin\imagecoffin
   }
   \group_end:
 }
\NewDocumentCommand{\setlabel}{m}
 {
  \keys_set:nn { miguel/label } { #1 }
 }

\cs_new_protected:Nn \miguel_includegraphics:nn
 {
  \includegraphics[#1]{#2}
 }
\cs_generate_variant:Nn \miguel_includegraphics:nn { V }

\ExplSyntaxOff

\begin{document}
\setlength{\parskip}{0pt}
\title{\vspace{-15mm}\fontsize{20pt}{10pt}\selectfont\textbf{Examining the EMC Effect using the \\$F_2^{n}$ neutron structure function}} 

\author{H. Szumila-Vance}\affiliation{Thomas Jefferson National Accelerator Facility, Newport News, VA}
\author{C.E. Keppel}\affiliation{Thomas Jefferson National Accelerator Facility, Newport News, VA}
\author{S. Escalante}\affiliation{Virginia Union University, Richmond, VA}
\author{N. Kalantarians}\affiliation{Virginia Union University, Richmond, VA}






\begin{abstract}
The persistently mysterious deviations from unity of the ratio of nuclear target structure functions to those of deuterium as measured in deep inelastic scattering (often termed the ``EMC Effect") have become the canonical observable for studies of nuclear medium modifications to free nucleon structure in the valence regime. The structure function of the free proton is well known from numerous experiments spanning decades. The free neutron structure function, however, has remained difficult to access. Recently it has been extracted in a systematic study of the global data within a parton distribution function extraction framework and is available from the CTEQ-Jefferson Lab (CJ) Collaboration. Here, we leverage the latter to introduce a new method to study the EMC Effect in nuclei by re-examining existing data in light of the the magnitude of the medium modifications to the free neutron and proton structure functions independently. From the extraction of the free neutron from world data, it is possible to examine the nuclear effects in deuterium and their contribution to our interpretation of the EMC Effect. In this study, we observe that the ratio of the deuteron to the sum of the free neutron and proton structure functions has some $x_{B}$ and $Q^{2}$ dependencies that impact the magnitude of the EMC Effect as typically observed. Specifically, different EMC slopes are obtained when data from different $x_{B}$ and $Q^{2}$ values are utilized. While a linear correlation persists between the EMC and short range correlation effects, the slope is modified when deuteron nuclear effects are removed.
\end{abstract}

\maketitle

\section{Introduction}

This work examines deep inelastic scattering data from the SLAC E139 experiment \cite{Gomez} to evaluate the contribution of the medium modification in the deuteron. Specifically, we use the recently extracted free neutron structure function, $F_2^n$, from the Jefferson Lab-CJ Collaboration to observe the deuteron medium modifications as they are included in the EMC effect slope. The EMC effect broadly describes the magnitude of nuclear medium modifications to bound nucleons for $x_B$ from 0.3-0.7, that is observed from the $F_2^A/F_2^d$ ratio. Significantly, we observe a $Q^2$ dependence in the SLAC data at large $x_B$ that is also observed in the free proton and neutron structure functions. This dependence impacts our extraction of the magnitudes and subsequent observations related to the EMC effect. 

The nuclear medium modifications in the deuteron can result from several effects to include off-shellness and properties of the wave function. While our observations are not the result of a precision study, we wish to point out specific considerations to bear in mind when studying the EMC effect. We observe at large $x_B$, particularly at 0.6 and greater: 
\begin{itemize}
    \item there exists $Q^2$ dependencies in the data for $F_2^A/F_2^d$
    \item the $Q^2$-dependence effect persists in the ratio of $F_2^A$ to the sum of the free proton and neutron structure functions and is different in magnitude
    \item these large modifications in the deuteron at large $x_B$ complicate the straight-forward extraction of universal observations of medium modifications in heavy nuclei, including short range correlation effects
\end{itemize}
Deep inelastic scattering (DIS) has produced a wealth of information on the partonic structure of nuclei spanning several decades of experiments. The EMC Effect, or the observed modification to the bound nucleon partonic structure as compared to that of the free nucleon, is a direct observation from these experiments. The EMC Effect is observed in the regime where $x_B$ is from 0.3--0.7, and the magnitude may be measured as the slope of the ratio of the $F_2^A/F_2^d$ nucleon structure function ratio where $A$ indicates the target nucleus and $d$ is deuterium. It was previously anticipated that this slope should be flat, but the surprising results indicated there is an $A$-dependent negative slope with increasing $x_B$. The interpretation of this modification has been the subject of many experiments and sustained theoretical efforts for several decades.

From DIS, a great deal of information is known about the proton, but the neutron's partonic structure is less understood. A recent effort by the CTEQ-Jefferson Lab Collaboration has reviewed the world DIS data and extracted free neutron structure functions for all kinematics~\cite{FutureCJpaper}~\cite{shujie}. Knowledge of the neutron structure function is critical, for instance, to constraining the (down quark) $d(x,Q^2)$ parton distribution function (PDF) and reducing uncertainties on other PDFs thereby via the momentum sum rule.

\begin{figure}[H]
 \centering
\begin{minipage}{0.49\textwidth}
\setlabel{pos=sw,fontsize=\scriptsize,labelbox}
 \xincludegraphics[width=\textwidth,label=(a),labelbox=false,fontsize=\Large]{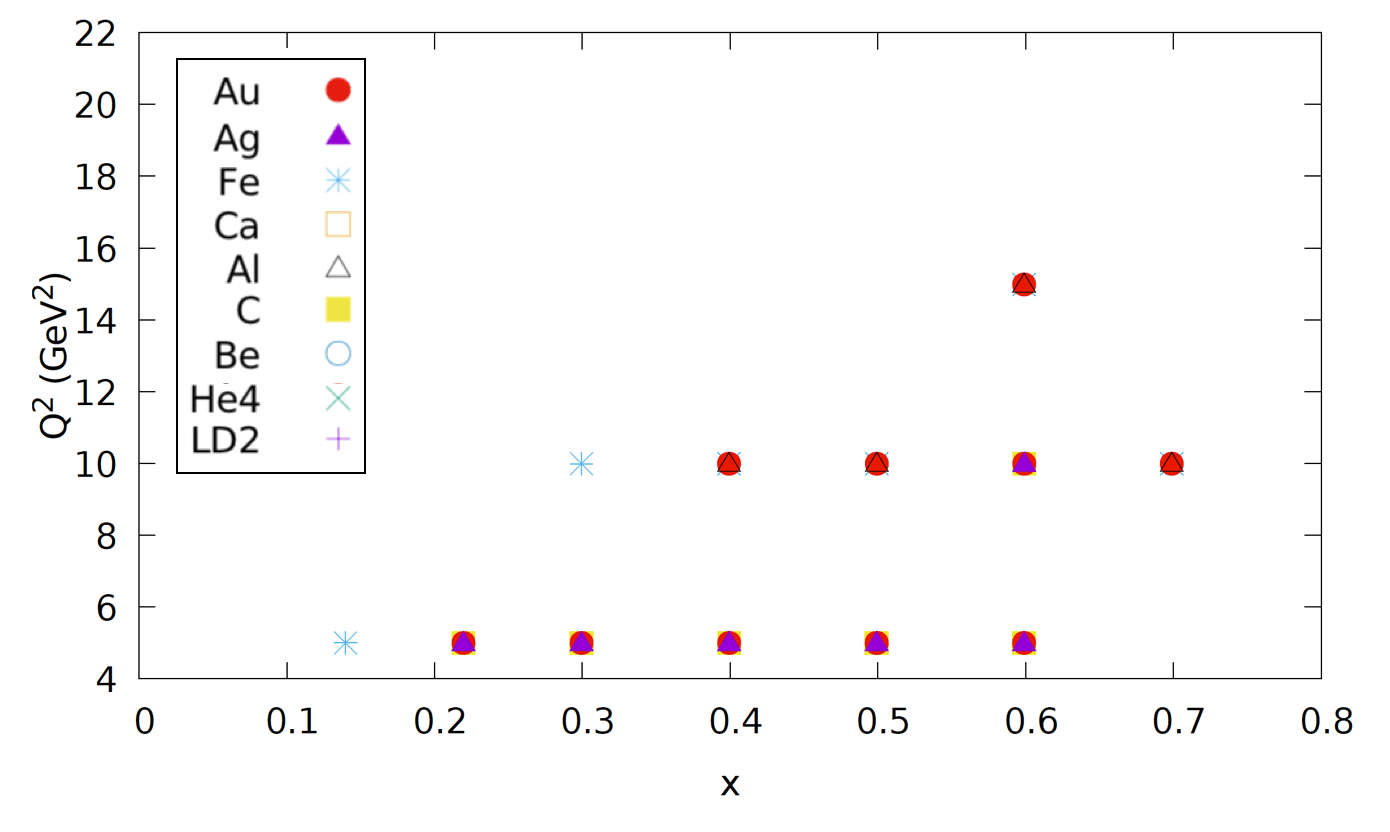}
\end{minipage}\hfill\begin{minipage}{0.49\textwidth}
\setlabel{pos=sw,fontsize=\scriptsize,labelbox}
 \xincludegraphics[width=\textwidth,label=(b),labelbox=false,fontsize=\Large]{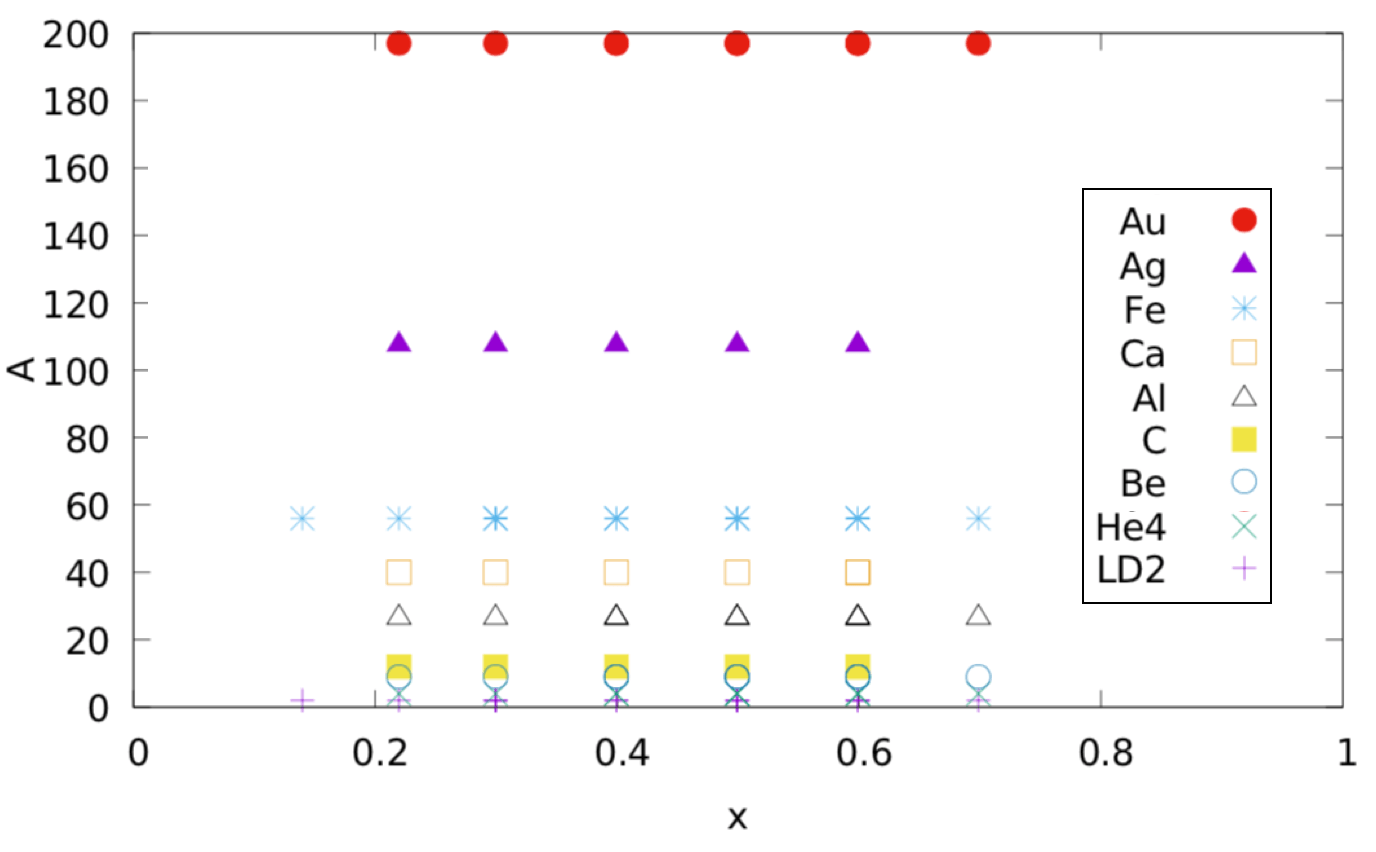}
 \end{minipage}
  \caption[]{(a) The kinematics of the SLAC E139 experiment are shown as they span $x_B$ and $Q^2$ for various target nuclei. (b) $x_B$ is shown per nuclei in the SLAC E139 experiment.}
  \label{fig:slac_q2x}
\end{figure}

Using the newly obtained free neutron structure function from CJ15, the EMC Effect can be examined in a different way. Rather than studying measured nuclear structure function ratios to deuterium, the effect on single nucleons investigated here was derived from measured nuclear and proton cross sections. The original SLAC E139 experiment~\cite{Gomez} is the only experiment to have published the measured cross sections in addition to the measured $F_2^A/F_2^d$ ratios. By extracting the $F_2^A$ structure function data starting from the cross sections of the SLAC E139 experiment, the EMC Effect can be understood in terms of the free proton and free neutron independently of the nuclear effects in deuterium. Fig.~\ref{fig:slac_q2x} shows the kinematics and targets of the E139 experiment in terms of the Bjorken scaling variable $x_B$ and the four-momentum transfer $Q^2$ in the inelastic scattering process.

\noindent The kinematics of the E139 experiment span the range of $x_B$ from approximately 0.1--0.7 and $Q^2$ from 2--15~GeV$^2$/c$^2$, depending on the target nucleus. The target nuclei include $^2$H, $^4$He, $^9$Be, $^{12}$C, $^{27}$Al, $^{40}$Ca, $^{56}$Fe, $^{108}$Ag, and $^{197}$Au. It is important to note that here, as with all of the fixed target lepton scattering EMC study data sets, $x_B$ and $Q^2$ are correlated variables. 
 
\section{Theory predictions using nuclear matter}

\begin{figure}[H]
  \centering
      	  \includegraphics[width=0.49\textwidth]{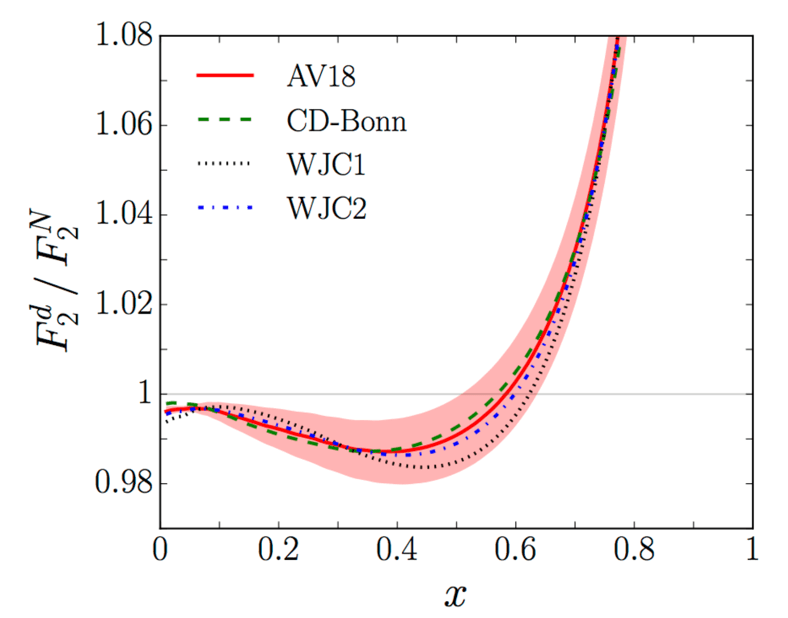}
 	 \caption[Theoretical extraction of the ratio of $F_2^d/F_2^N$]{The theoretical extraction of the ratio of $F_2^d/F_2^N$ where $N$ indicates per nucleon. The deuteron exhibits some $x_B$-dependence such that the ratio is modified by as much as 6$\%$ in the regime of $x_B$ between 0.3--0.7 of interest to the EMC Effect~\cite{cjSite}.}
  \label{fig:dn_theory}
 \end{figure}

Recent theoretical extractions of the $F_2^d$ structure function per nucleon are shown in Fig.~\ref{fig:dn_theory} from CJ15~\cite{CJ15Fit}. These results indicate that, for $x_B$ relevant to the EMC Effect (0.3--0.7), theoretical models show a nuclear modification of nearly $-2\%$ at $x_B<0.5$, increasing and changing sign up to $+4\%$ at $x_B\approx0.75$. While this may not seem large, we note that the EMC Effect at its largest is about a $15\%$ effect in total. Therefore, the deuteron nuclear effects alone may account for up to about $30\%$ of the observed nuclear medium modifications. This general shape is common with other extractions, all predicting a steep rise in deuteron nuclear effects at the largest $x_B$, although the CJ rise shows an onset at somewhat lower $x_B$ than some. Importantly, the nuclear effects in deuterium do not exhibit a linear behavior in the $0.3<x_B<0.7$ region typically studied for the EMC Effect using $F_2^d$ in the denominator. Another common feature to deuteron nuclear effect theory is a predicted $Q^2$ dependence (a feature also shared, for instance, with predictions by Kulagin-Petti~\cite{AKP_2017} albeit at somewhat different magnitudes). We note here again that, for most of the EMC-type data sets, $x_B$ and $Q^2$ are correlated variables. 

\begin{figure}[H]
  \centering
      	  \includegraphics[width=0.49\textwidth]{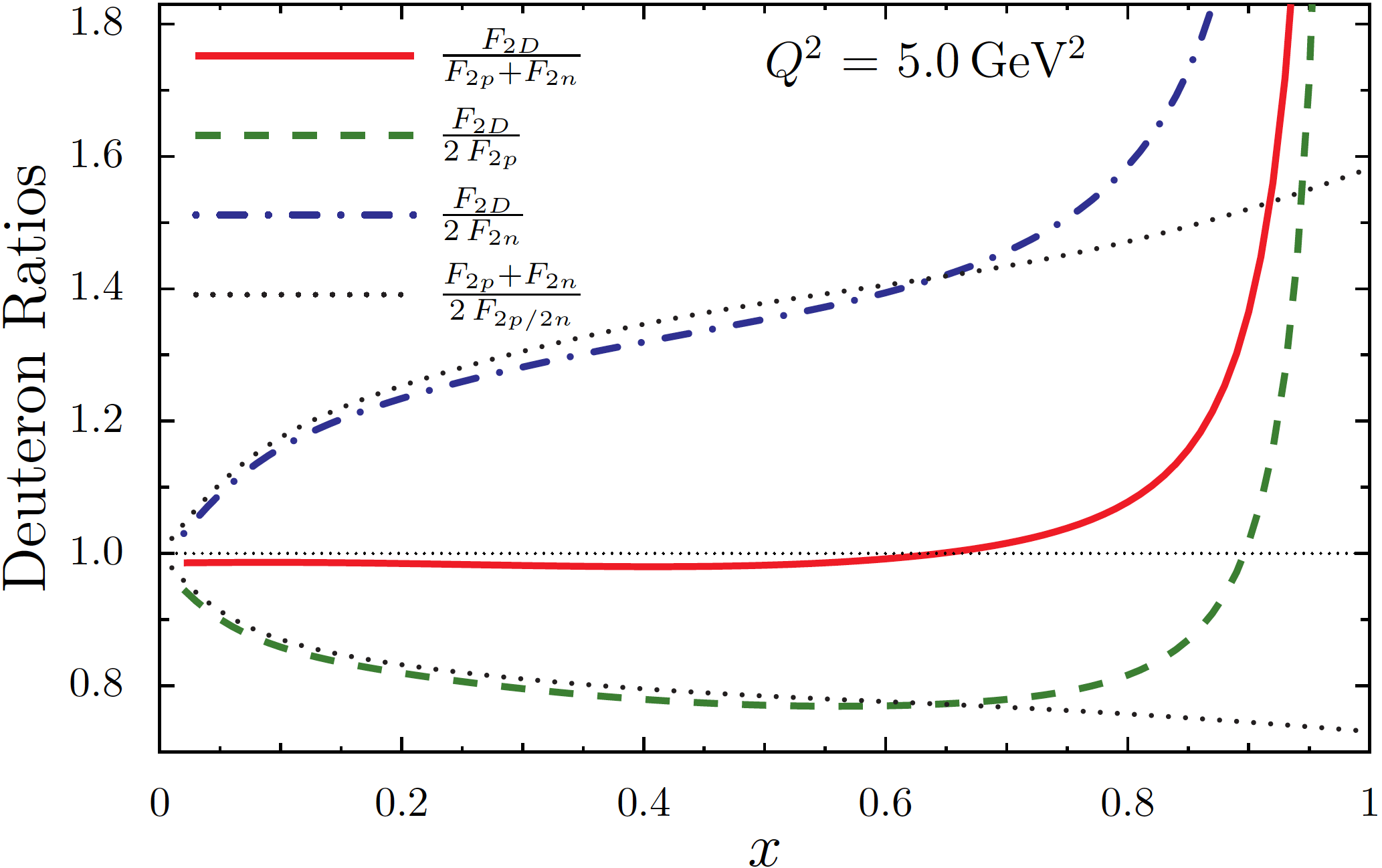}
 	 \caption[Theoretically-derived deuteron $F_2$ ratios with respect to the free nucleons]{Theoretically-derived deuteron $F_2$ ratios with respect to the free neutron and proton. These results are shown for fixed $Q^2=5$~GeV$^2$/c$^2$.}
  \label{fig:deut_theory}
 \end{figure}

Beginning with quark models and choosing $Q^2=5$~GeV$^2$/c$^2$, theoretically-derived $F_2$ deuterium ratios with respect to the free neutron and free proton from Cloet~\cite{Cloet:2005pp,Cloet:2006bq,Cloet:2012td} are shown in Fig.~\ref{fig:deut_theory}~\cite{cloet}. 
  \noindent Here, the $F_2$ deuterium ratio with respect to the sum of the free neutron and free proton contributions is shown in red dipping just below unity in the EMC Effect regime and rising at large $x_B$, as CJ qualitatively describes. By taking the ratio of $F_2^d/2F_2^p$ (shown in green dashed) where $p$ indicates the proton structure function, the slope dips well below unity and is far greater in magnitude than the $F_2$ deuterium ratio shown in red (solid). Taking $F_2^d/2F_2^n,$ where $n$ indicates the neutron structure function, shows conversely a positive slope that is of greater magnitude still than $F_2^d/2F_2^p$. The proton dominance is expected, given that the $d(x_B)/u(x_B)$ (quark) PDF is predicted to decline substantially at large $x_B$ and so similarly $F_2^n/F_2^p$. It is nonetheless interesting that the size of the nuclear effects is predicted to be roughly equal to the conventional $F_2^A/F_2^d$ EMC Effect alone for the proton, but twice this EMC Effect magnitude for the neutron. 

\section{$F_2^n$ extraction and the CJ15 fit}
The CTEQ-Jefferson Lab (CJ) Collaboration recently reviewed the world DIS data on the proton and deuteron structure functions, including all inclusive data from the Jefferson Lab 6 GeV program and performed an extraction of the free neutron structure function from proton and deuteron data over the full available kinematics. The extraction utilized the CJ15 simultaneous QCD fit of PDFs and deuteron nuclear corrections \cite{CJ15_paper} to calculate the $d/(p+n)$ ratio and convert experimental data on the deuteron and proton obtained at matching kinematics to neutron data. The uncertainties of the extracted neutron structure function include the choice of wave function in the nucleon smearing, off-shell ansatz, higher twist form, target mass approach, form for parton distribution behavior at large $x$, and multi data set normalizations\cite{FutureCJpaper}. In this paper, we utilize this extracted free neutron structure function data to examine the EMC Effect in a new way. 

 \begin{figure}[H]
 \centering
\begin{minipage}{0.48\textwidth}
\setlabel{pos=sww,fontsize=\scriptsize,labelbox}
 \xincludegraphics[width=\textwidth,label=(a),labelbox=false,fontsize=\Large]{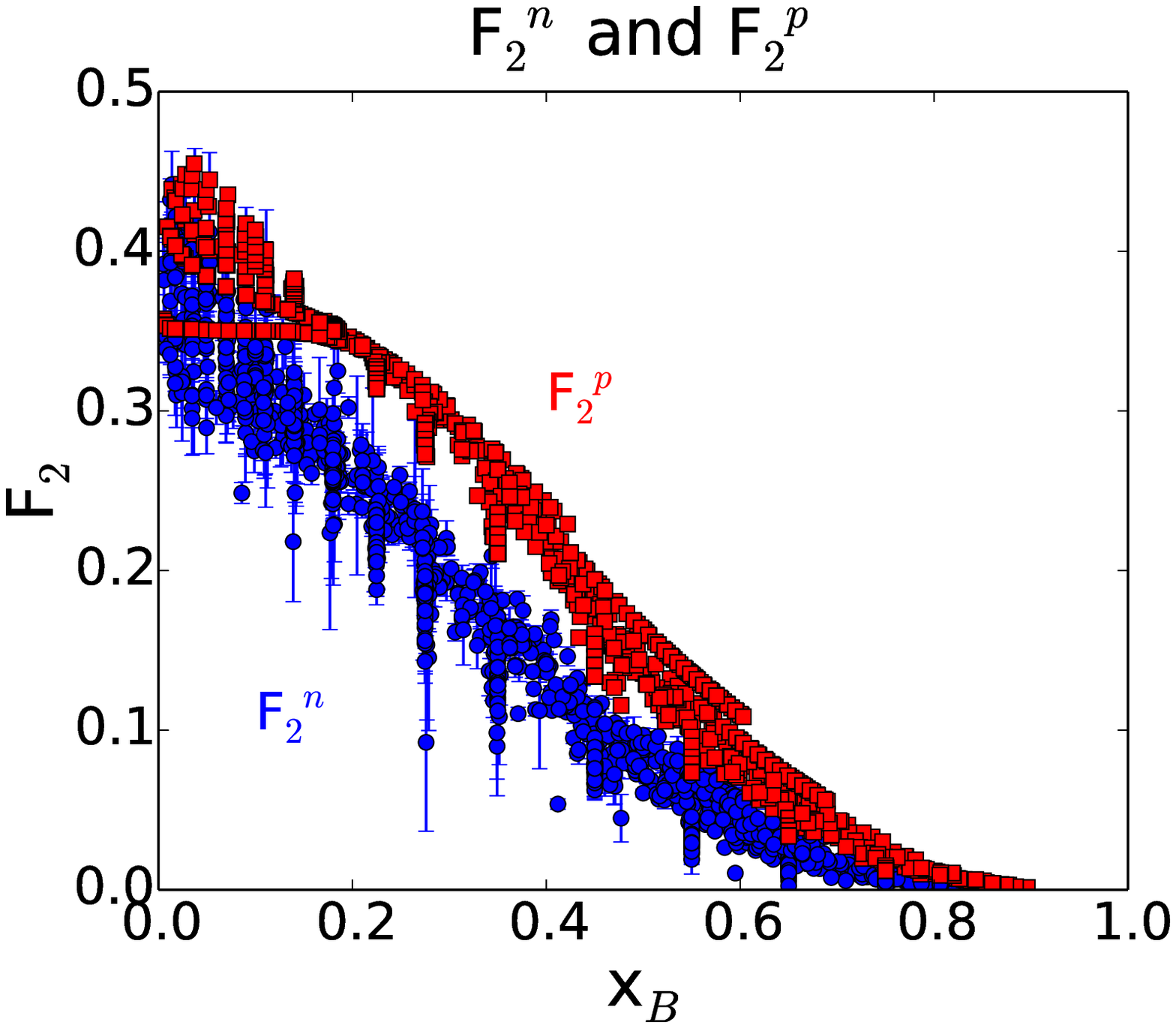}
\end{minipage}\hfill\begin{minipage}{0.48\textwidth}
\setlabel{pos=sww,fontsize=\scriptsize,labelbox}
 \xincludegraphics[width=\textwidth,label=(b),labelbox=false,fontsize=\Large]{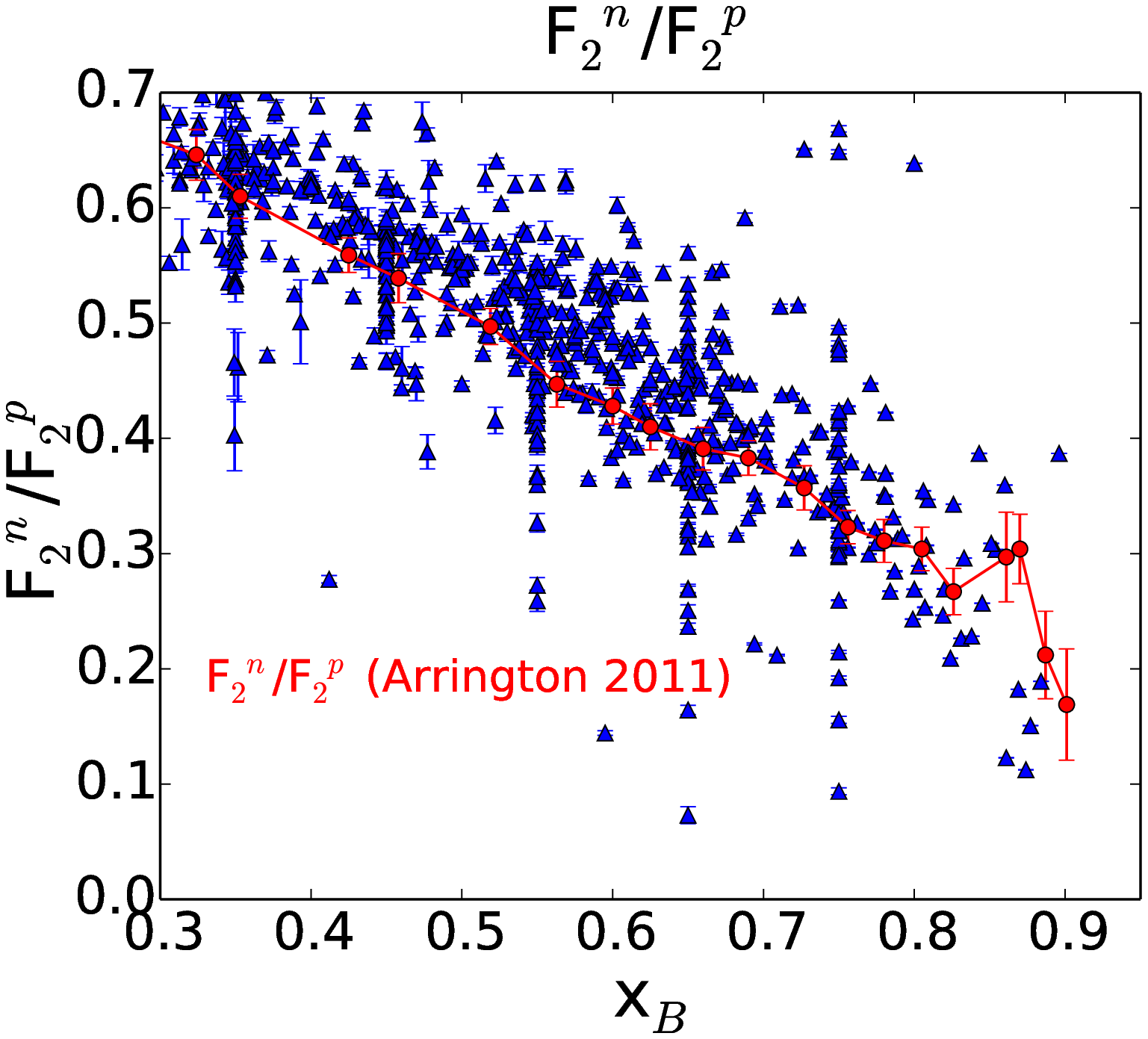}
 \end{minipage}
 \caption[]{(a) The extracted CJ $F_2^n$ DIS data are shown in blue circles, and the $F_2^p$ from the NMC global fit is shown in red squares for the same corresponding $x_B$ and $Q^2$~\cite{NMC}. (b) The ratio $F_2^n/F_2^p$ from the (a) is shown as a function of $x_B$ in the blue triangles. A comparison to a previous extraction of the $F_2^n/F_2^p$ ratio from \cite{Arrington} is shown by the red circles and line.}
  \label{fig:F2np_general}
\end{figure} 
One should note that a QCD-based fit also aiming to include the valence regime, and therefore somewhat similar to the CJ15 fit, was published in 2017 by Alekhin, Kulagin and Petti (AKP) \cite{AKP_2017}. This resulted in a markedly different fitted offshell correction function, the reasons for which have been actively investigated by both collaborations \cite{accardi:dnp2020} and will be incorporated in part into theoretical uncertainties.The AKP $F_2^n$ structure function is, roughly speaking, about 10\% different than its CJ15 counterpart, crossing at $x\approx0.4$ and displaying a steeper slope. However, information was not readily available for the resultant $F_2^n$ structure function, and so we cannot quantify the differences this would induce on the results we present here. Importantly, we do expect the general observations made here to be independent of the global fit choice. 


In Fig.~\ref{fig:F2np_general}, general comparisons of the newly extracted $F_2^n$ are made with respect to $F_2^p$ obtained from an NMC fit~\cite{NMC}. On the top of Fig.~\ref{fig:F2np_general}, the $F_2^n$ and $F_2^p$ for the same $x_B$ and $Q^2$ kinematics is shown. The $F_2^n$ in blue circles is extracted from the world data, and the $F_2^p$ is the corresponding value (for the same kinematics) from the NMC fit~\cite{NMC}. The overall $x_B$ dependence is somewhat different for the neutron and proton; the neutron structure function flattens earlier at larger $x_B$ while the proton structure function drops off more sharply. The ratio of these quantities is shown in Fig.~\ref{fig:F2np_general}(b) and agrees with previously extracted results~\cite{Arrington}. The bands and striping reflect $Q^2$ differences at fixed $x_B$.
 
\subsection{$Q^2$ dependence}

\begin{figure}[H]
\begin{minipage}{0.44\textwidth}
\setlabel{pos=sww,fontsize=\scriptsize,labelbox}
 \xincludegraphics[width=\textwidth,label=(a),labelbox=false,fontsize=\Large]{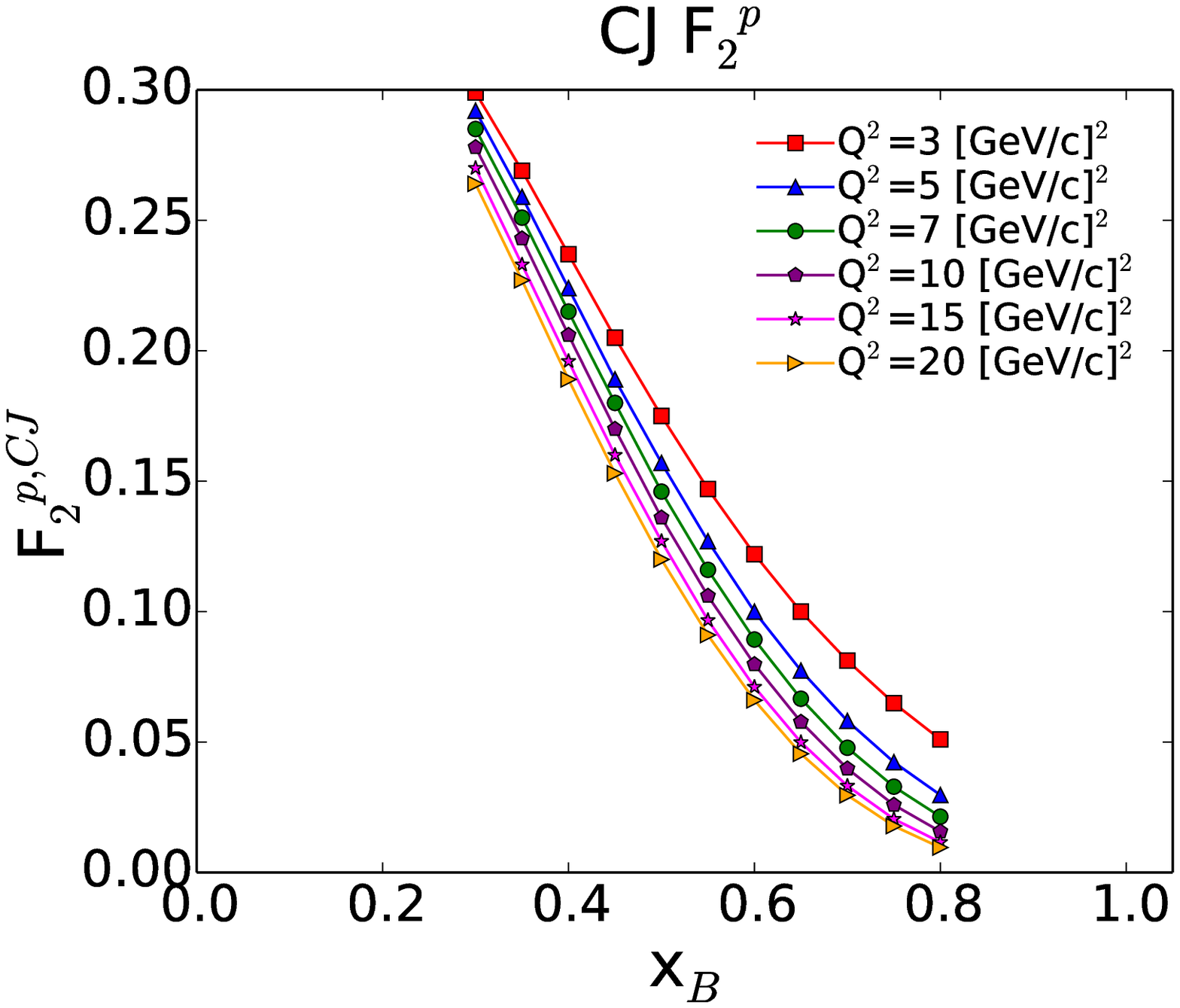}
\end{minipage}\hfill\begin{minipage}{0.44\textwidth}
\setlabel{pos=sww,fontsize=\scriptsize,labelbox}
 \xincludegraphics[width=\textwidth,label=(b),labelbox=false,fontsize=\Large]{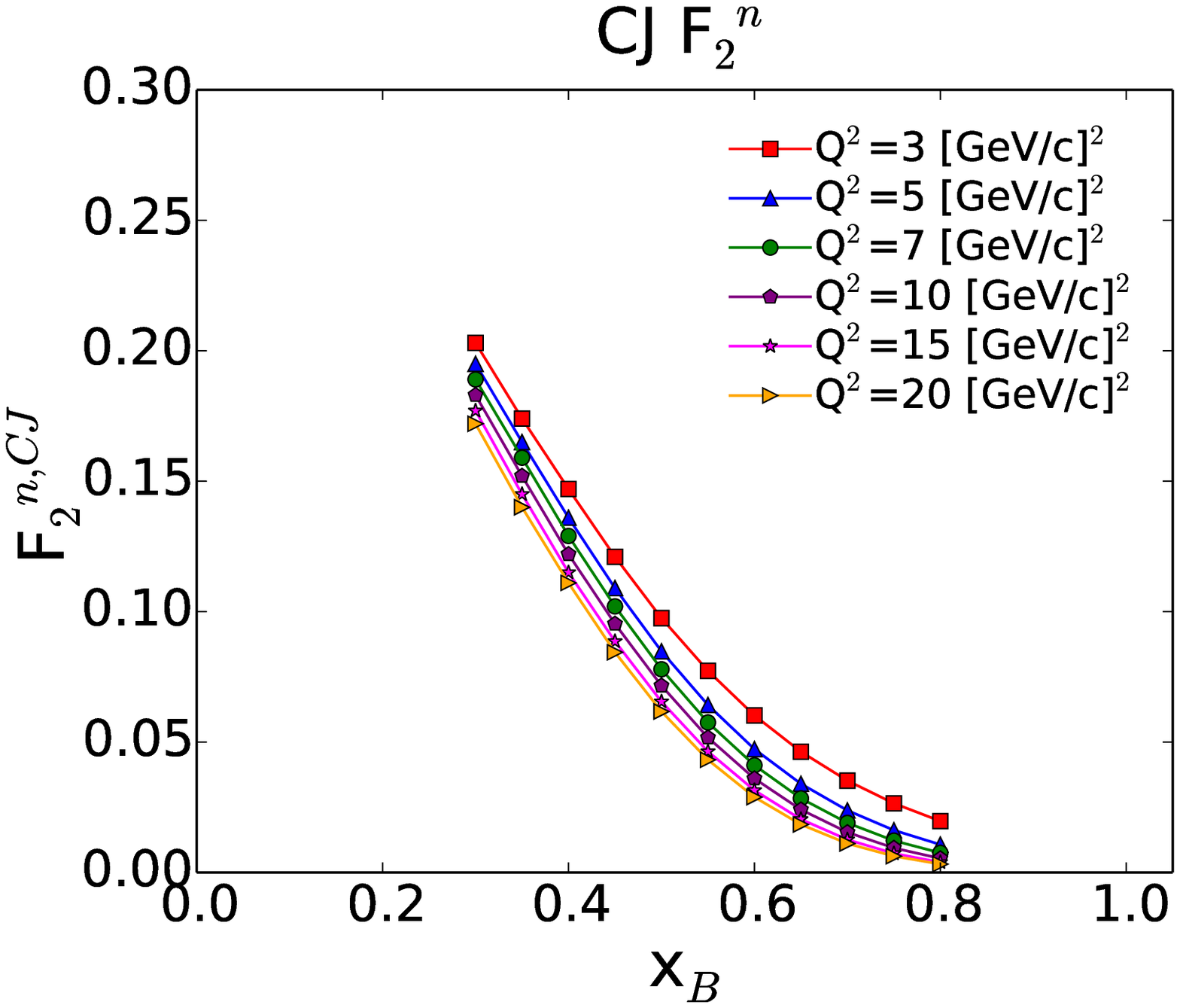}
 \end{minipage}\hfill\begin{minipage}{0.44\textwidth}
 \setlabel{pos=sww,fontsize=\scriptsize,labelbox}
 \xincludegraphics[width=\textwidth,label=(c),labelbox=false,fontsize=\Large]{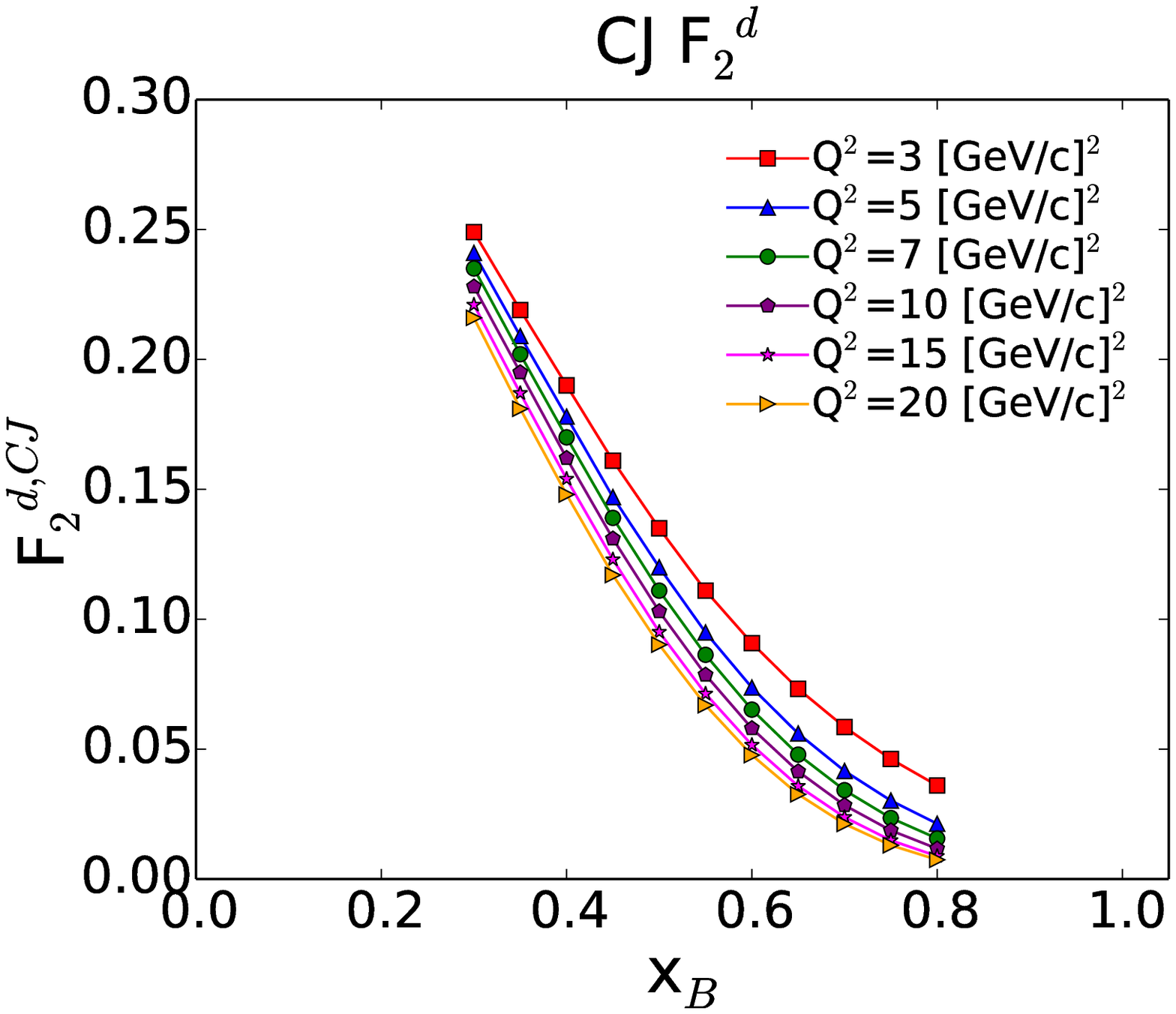}
 \end{minipage}
  \caption[Proton and deuteron from CJ15]{(a) The proton structure function from CJ15 is shown as a function of $x_B$ for various fixed $Q^2$ (as indicated). (b) The neutron structure function from CJ15 is shown as a function of $x_B$ for various fixed $Q^2$. (c) The deuteron structure function from CJ15 is shown as a function of $x_B$ for various fixed $Q^2$. All are shown without uncertainty to better reflect the $Q^2$ dependence of each.}
  \label{fig:pd_CJ15}
\end{figure}

A significant observation in the CJ15 extracted structure functions is the $Q^2$ dependence. The proton structure function is shown in Fig.~\ref{fig:pd_CJ15}(a) for various $Q^2$ from the CJ15 global PDF fit. The middle plot shows the same $Q^2$ dependence for the neutron structure function from CJ, and the corresponding deuteron structure function is shown in Fig.~\ref{fig:pd_CJ15}(c). Here, no uncertainties are shown as we focus on the $Q^2$ dependence. The SLAC E139 experiment took measurements at $Q^2$ in the range of 5--15~GeV$^2$/c$^2$. The nucleon and deuteron structure functions from CJ show a clear $Q^2$ dependence for fixed values of $x_B$ in this range of $Q^2$. This can translate into a $Q^2$ dependence in the EMC Effect ratio $F_2^A/F_2^d$ if the deuteron $Q^2$ dependence is different from the heavier nuclei and ultimately contributes to understanding medium modifications to the free nucleon taking into account proton and neutron individual differences.

\begin{figure}[H]
\begin{minipage}{0.48\textwidth}
\setlabel{pos=sww,fontsize=\scriptsize,labelbox}
 \xincludegraphics[width=\textwidth,label=(a),labelbox=false,fontsize=\Large]{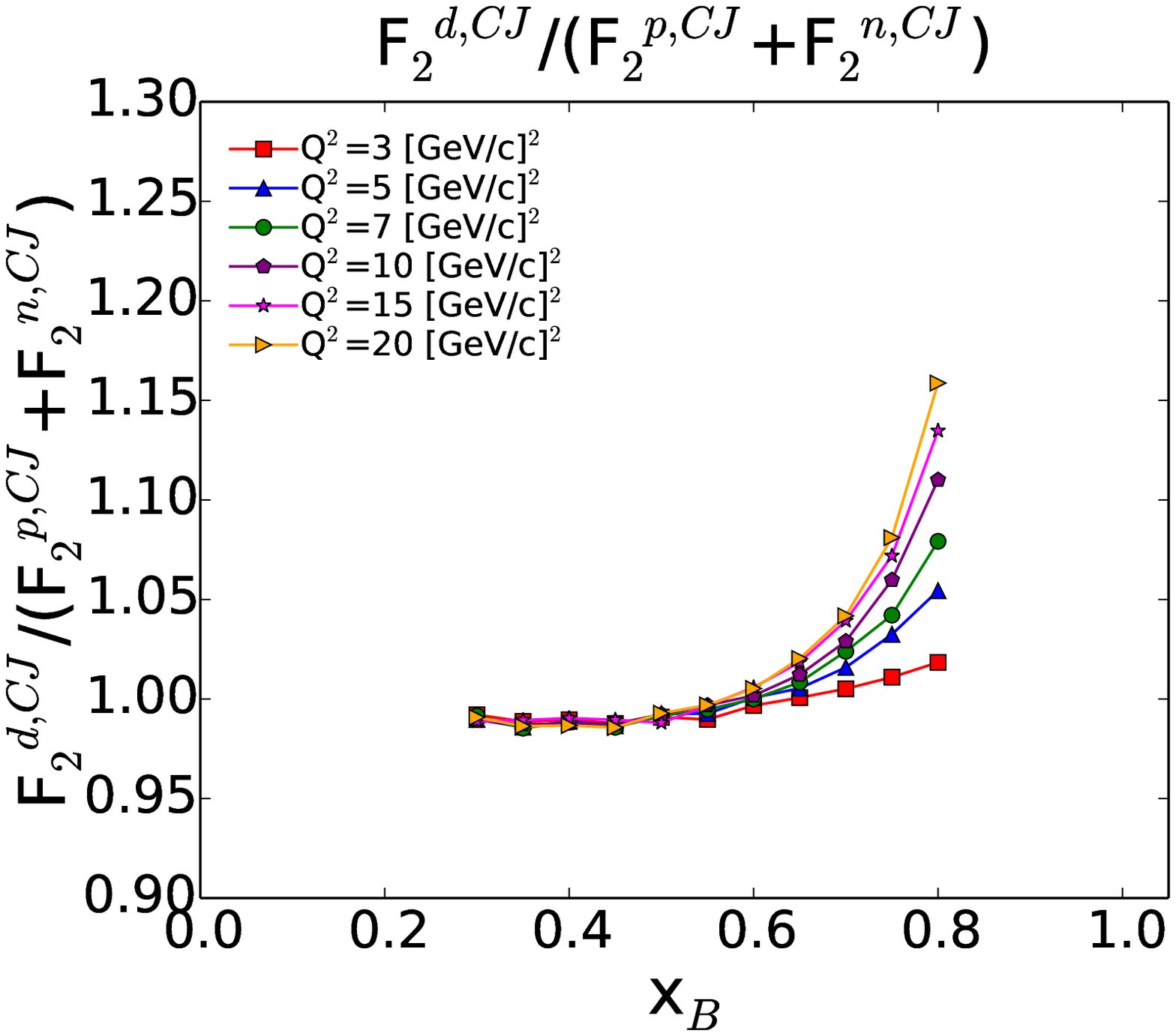}
\end{minipage}\hfill\begin{minipage}{0.48\textwidth}
\setlabel{pos=sww,fontsize=\scriptsize,labelbox}
 \xincludegraphics[width=\textwidth,label=(b),labelbox=false,fontsize=\Large]{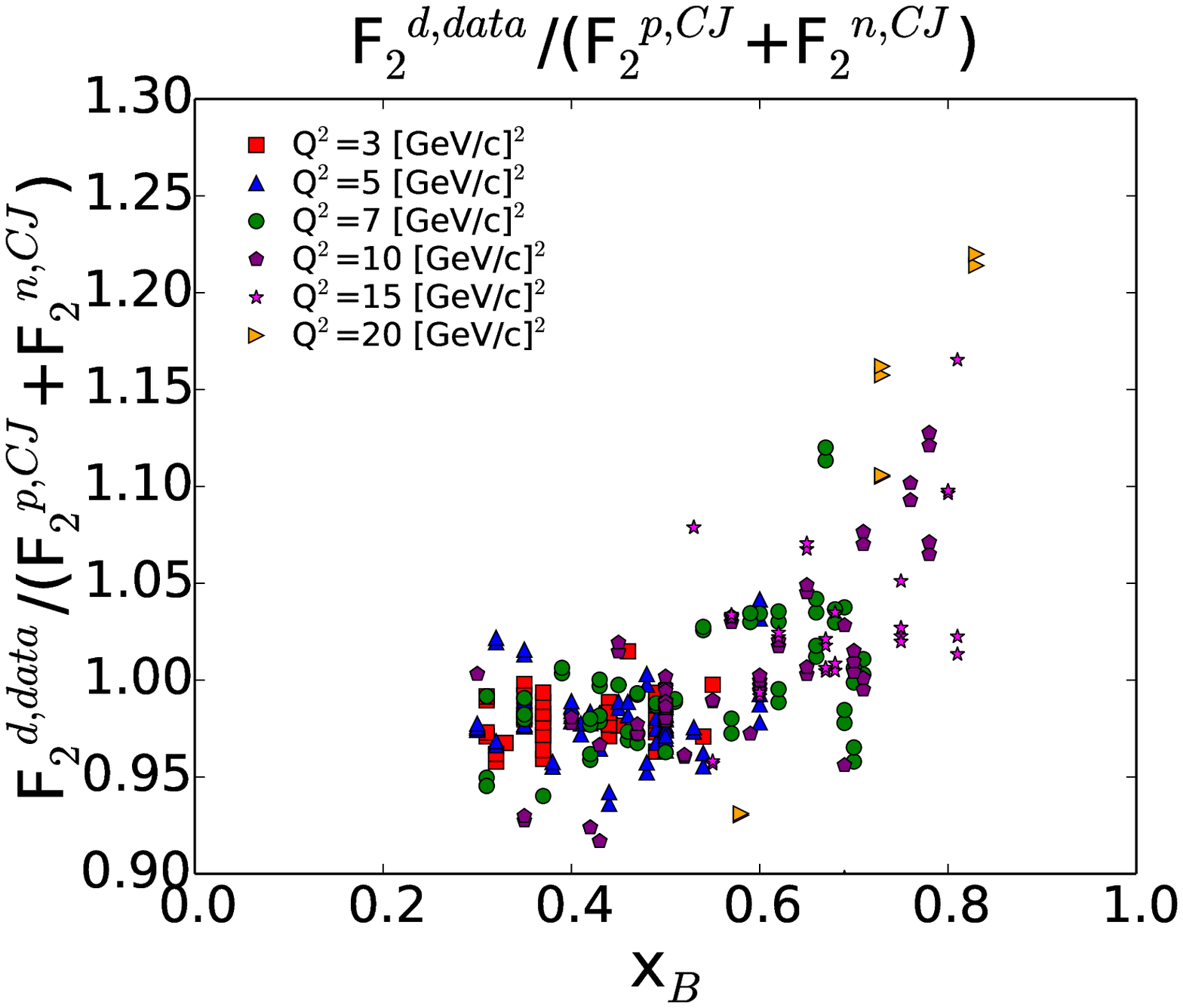}
 \end{minipage}
  \caption[]{(a) The deuteron structure function divided by the sum of the free proton and free neutron structure functions from the CJ15 fit is shown for various $Q^2$. This ratio reflects the magnitude of the nuclear effects in the deuteron. The $Q^2$ dependence shows significant spread above $x_B=0.6$ where the ratio begins to increase. (b) The Whitlow deuterium data from SLAC~\cite{XS_d} is shown divided by the sum of the free proton and free neutron structure functions from the CJ15 fit. In both cases, uncertainties are not shown in order to emphasize the general kinematic dependence.}
  \label{fig:dpn_cj}
\end{figure}

The nuclear effects of the proton and neutron in the deuteron can be directly characterized by dividing the deuteron structure function by the sum of the free proton and free neutron structure functions. The result is shown in Fig.~\ref{fig:dpn_cj}(a). Over most of the typical range of $x_B$ that is relevant to the EMC Effect, the $Q^2$ dependence is minimal. At approximately $x_B=0.6$ and greater, however, the ratio diverges greatly with $Q^2$ dependence.  

\begin{figure}[H]
\begin{minipage}{0.49\textwidth}
\setlabel{pos=sww,fontsize=\scriptsize,labelbox}
 \xincludegraphics[width=\textwidth,label=(a),labelbox=false,fontsize=\Large]{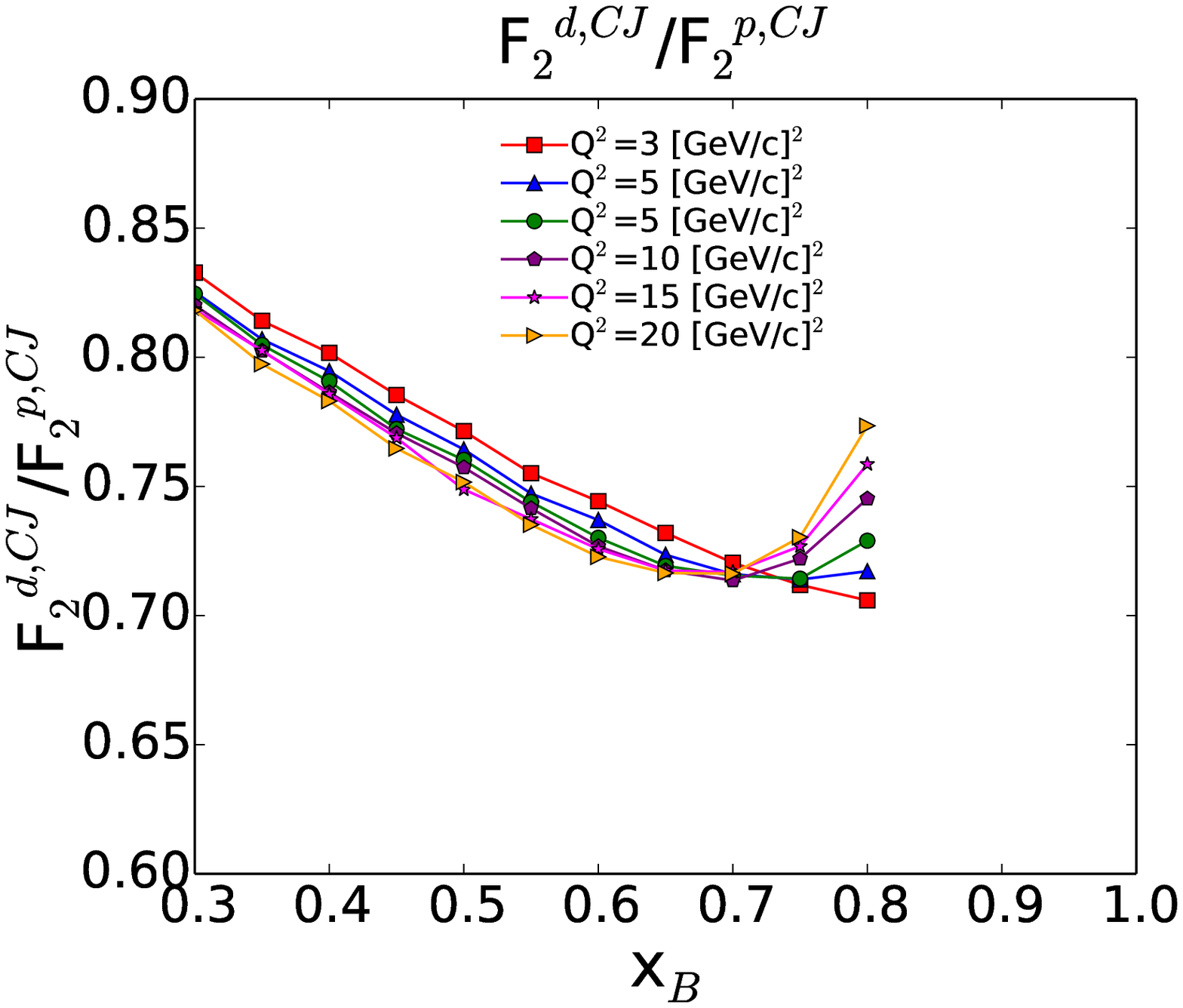}
\end{minipage}\hfill\begin{minipage}{0.49\textwidth}
\setlabel{pos=sww,fontsize=\scriptsize,labelbox}
 \xincludegraphics[width=\textwidth,label=(b),labelbox=false,fontsize=\Large]{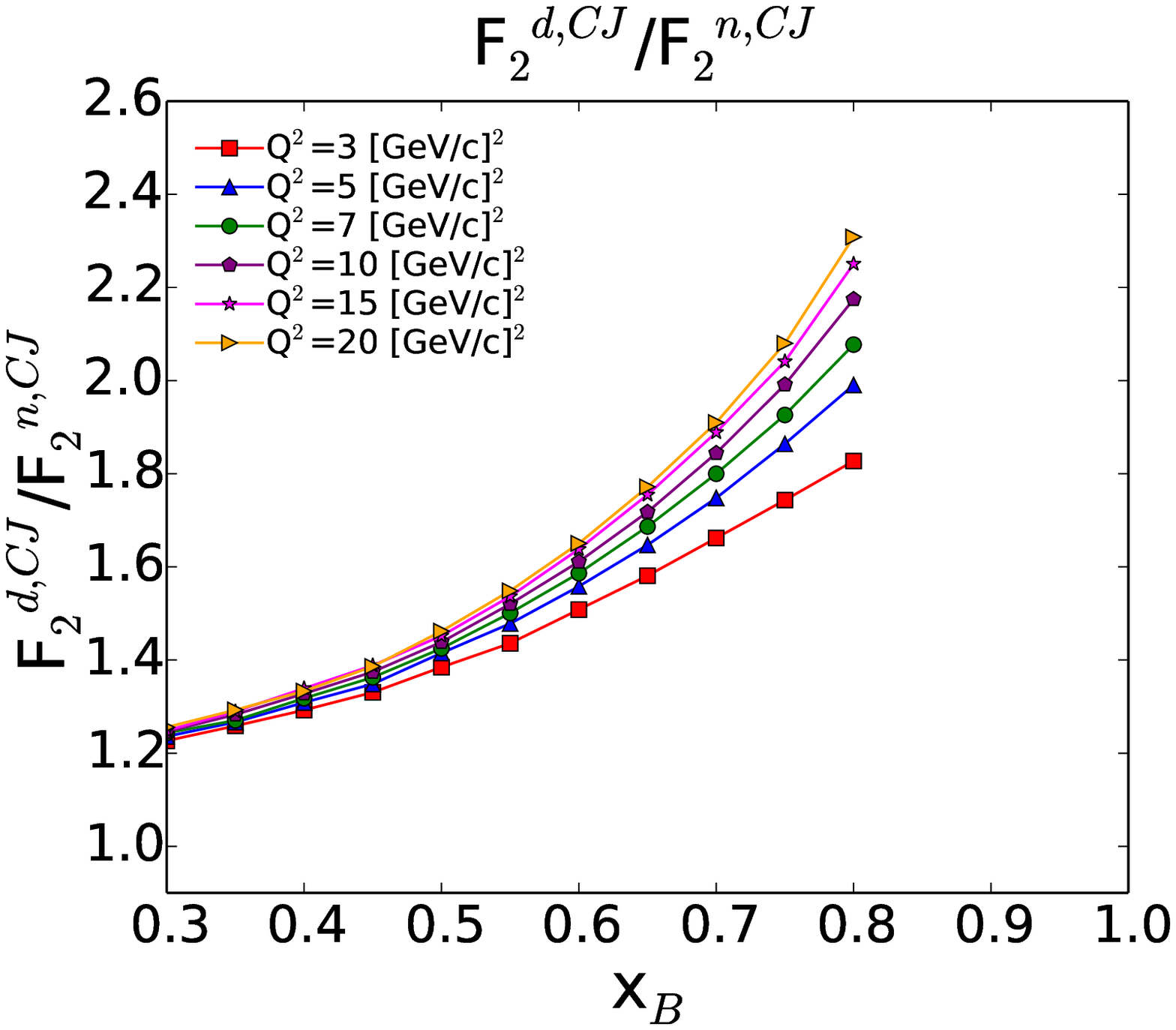}
 \end{minipage}
  \caption[Deuteron ratios from CJ15]{(a) The deuterium structure function from the CJ15 fit is shown as a ratio to the proton structure function from CJ15 for various $Q^2$. (b) The deuterium structure function from the CJ15 fit is shown as a ratio to the neutron structure function from CJ15 for various $Q^2$. Uncertainties are not shown to emphasize the kinematic dependence of the structure functions.}
  \label{fig:dratio_cj}
\end{figure}

The same ratio of deuterium to the sum of free nucleons is shown in Fig.~\ref{fig:dpn_cj}(b) using SLAC deuterium data~\cite{Whitlow,whitlow_pl}. The same general trend is observed as that observed in CJ15 fits where $Q^2$ has a more significant effect on the ratio at large values of $x_B$. The separate $F_2^d/F_2^p$ and $F_2^d/F_2^n$ ratios from the CJ15 fit are shown in Fig.~\ref{fig:dratio_cj}. Here, it is useful to note that $F_2^d/F_2^p$, which is independent of theoretical bias, shows large $x_B$ and $Q^2$ dependence.  
 
In Fig.~\ref{fig:dratio_cj}, the deuterium ratio to the neutron structure function appears to have a general divergence that increases with increasing $x_B$ while the ratio to the proton has a more consistent offset throughout the range of $x_B<0.7$ relevant to EMC studies. {\it It is important to note the magnitude of this spread in ratios at large $x_B$ values.} This effect has likely gone relatively unnoticed prior to this work because many experiments, SLAC E139 included, averaged $Q^2$ to achieve statistical precision for each point in $x_B$~\cite{Gomez}. 

The physical origin for the observed $Q^2$ dependence is unknown. Potential isospin dependence~\cite{TThomas}, and in particular isospin dependence of higher twist terms has previously been investigated~\cite{Liuti, AKP_2017}. The higher twist was found to be consistent with zero within the errors, with some deviation from zero roughly above $x = 0.7$, where the neutron and proton were found to differ somewhat. However, the uncertainties, including radiative corrections, are large in this region. 

Most fits and models characterizing the EMC Effect (for instance, \cite{Malace}) often do not include any $Q^2$ dependence. The two observations presented here, i.e. the $x_B$ and the $Q^2$ dependence of the deuterium nuclear effects, should be taken into account in the interpretation and extraction of the EMC Effect. For example, the EMC Effect has recently been characterized by linear fits in the range of $x_B$ from 0.3--0.7 where deuterium in fact exhibits {\bf non-linear} nuclear modification effects. 

\section{Structure function extraction from the E139 cross sections}

For this analysis, experimental cross sections (as opposed to published ratios) were desirable for the flexibility to construct and study ratios of nuclei to free nucleon quantities. The SLAC E139 experiment published the experimental cross sections for the following nuclei: $^2$H, $^4$He, $^9$Be, $^{12}$C, $^{27}$Al, $^{40}$Ca, $^{56}$Fe, $^{108}$Ag, and $^{197}$Au. The cross sections include published statistical and systematic errors. The $F_2$ structure function for each nucleus is extracted from the cross section using the relationship shown in Equation~\eqref{eq:f2eqn}:

\begin{equation}
F_2(x_B,Q^2) = \dfrac{d^2\sigma}{d\Omega dE'}\dfrac{1+R}{1+\epsilon R}\dfrac{K\nu}{4\pi^2\alpha\Gamma (1+\nu^2/Q^2)}
\label{eq:f2eqn}
\end{equation}

In Equation~\eqref{eq:f2eqn}, $\epsilon$, $K$, and $\Gamma$ are defined by the electron-nucleon deep inelastic scattering kinematics in Equation~\eqref{eq:f2eqndetails}. These quantities include the measured scattering angle $\theta$, the momentum transfer $Q^2$, the energy transfer $\nu$, the scattered electron energy $E'$, the mass of the proton $M$, and the squared invariant mass $W^2$. 

$R$, the longitudinal to transverse cross section ratio, in this analysis was determined from the R1990 fit~\cite{whitlow_pl} to world data, which assumes this quantity to be $A$-independent. It is critical to note that this may not be an accurate assumption. Recent results, measured at low $Q^2$ where the quantity is large enough to observe differences, display a difference in the deuteron and proton longitudinal cross sections~\cite{Tvaskis, Ioana}. This effect will be the subject of further experimental studies~\cite{PhysRevC.86.045201}~\cite{E12_14_002}.  

\begin{align}
\epsilon=&(1+2\dfrac{\nu^2+Q^2}{Q^2}tan^2\dfrac{\theta}{2})^{-1}\\
K=&\dfrac{W^2-M^2}{2M}\\
\Gamma=&\dfrac{\alpha KE'}{2\pi^2Q^2E(1-\epsilon)}
\label{eq:f2eqndetails}
\end{align}

For most published cross section ratios, corrections are applied to account for the excess of neutrons in asymmetric nuclei for comparison to deuterium. This correction is referred to as the ``isoscalar correction"and is defined in Equation~\eqref{eq:isocorr}. While this correction is not applied in this paper for the $F_2$ ratio results per free proton and free neutron, it is relevant for accurately reconstructing the published $F_2^A$/$F_2^d$ ratios in most analyses. For analysis in this paper, an additional requirement for $W^2>4$~GeV$^2$/c$^2$ was applied in order to ensure that the results obtained are above the resonance region.  

\begin{equation}
f_{iso}^A = \dfrac{\dfrac{1}{2}(1+F_2^n/F_2^p)}{\dfrac{1}{A}(Z+(A-Z)F_2^n/F_2^p)}
\label{eq:isocorr}
\end{equation}
\noindent 

\section{EMC studies per nucleon}

\begin{figure}[H]
\begin{minipage}{0.49\textwidth}
\setlabel{pos=sww,fontsize=\scriptsize,labelbox}
 \xincludegraphics[width=\textwidth,label=(a),labelbox=false,fontsize=\Large]{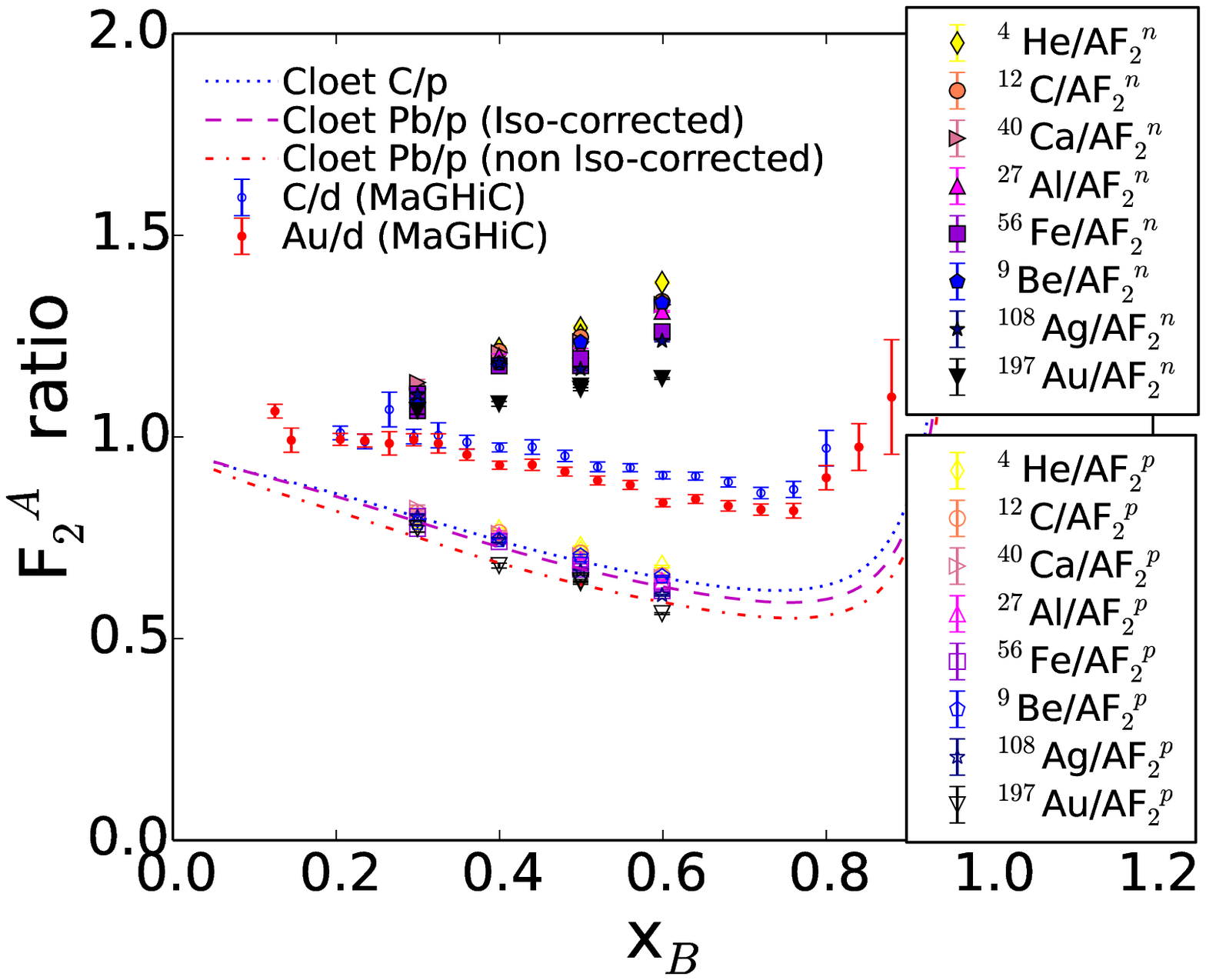}
\end{minipage}\hfill\begin{minipage}{0.49\textwidth}
\setlabel{pos=sww,fontsize=\scriptsize,labelbox}
 \xincludegraphics[width=\textwidth,label=(b),labelbox=false,fontsize=\Large]{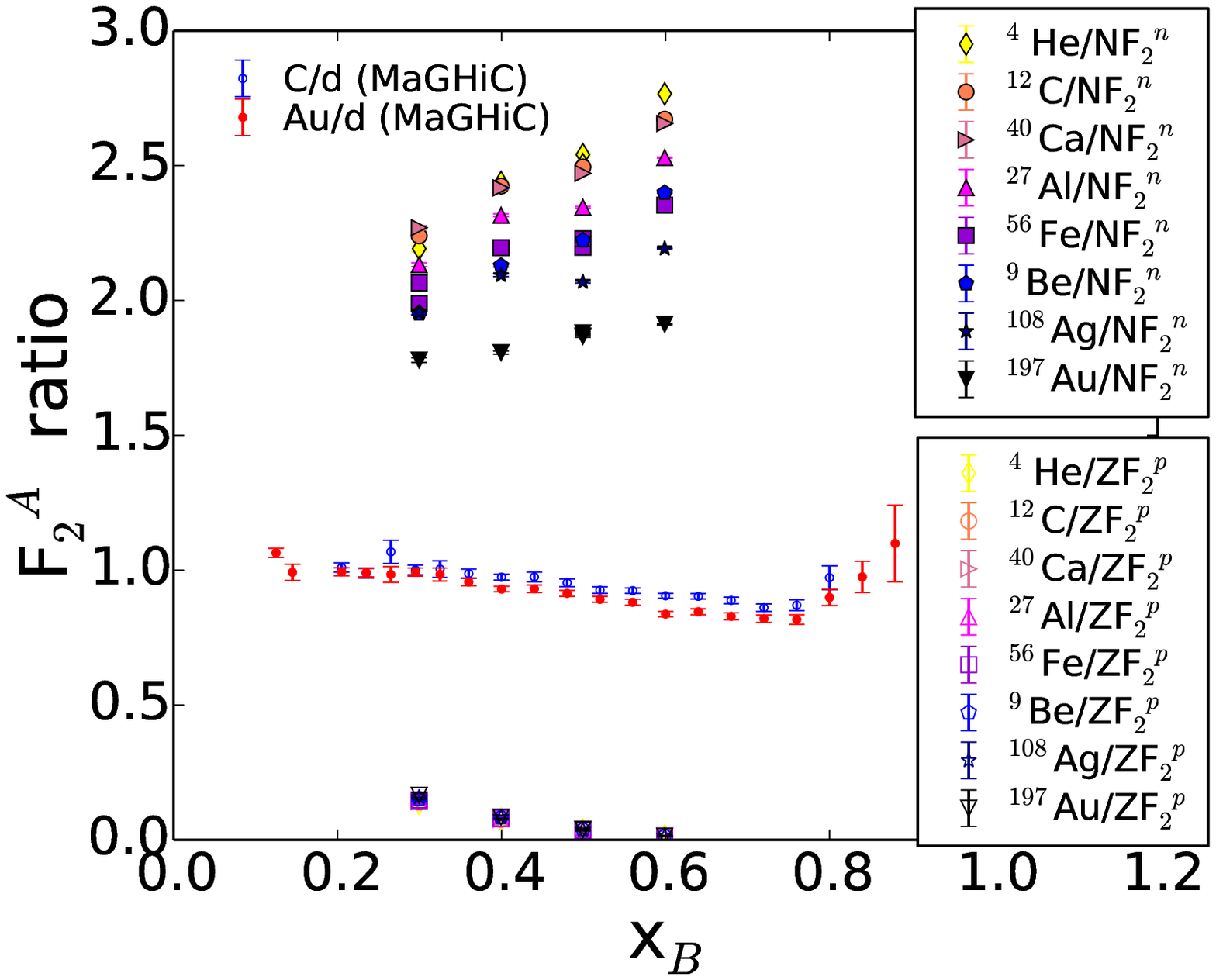}
 \end{minipage}
  \caption[$F_2^A$ ratio to $F_2^n$ and $F_2^p$]{(a) $F_2^A$ calculated from the published SLAC E139 cross sections is taken as a ratio per nucleon to $F_2^n$ and $F_2^p$, separately. The published ratios for carbon and gold~\cite{Malace} are shown for reference as $F_2^A/F_2^d$ per nucleon. Theory predictions for the $F_2^A$ structure function per nucleon as a ratio to $F_2^p$ are also shown~\cite{cloet}. The nuclei from the SLAC E139 experiment are shown as ratios per nucleon relative to $F_2^p$ (bottom) and $F_2^n$ (top). (b) $F_2^A$ calculated from the published SLAC E139 cross sections is taken as a ratio per neutron or proton and $F_2^n$ and $F_2^p$, separately. The published EMC ratios for carbon and gold~\cite{Malace} are shown for reference. Different from the plot on the left (a), the nuclei from the SLAC E139 experiment are shown as ratios per proton relative to $F_2^p$ (bottom) and per neutron $F_2^n$ (top). A larger spread in the nuclei is observed per neutron that per proton for $x_B$. }
  \label{fig:data_np_ratio}
\end{figure}

The standard EMC structure function ratio of nucleus to deuterium ($F_2^A/F_2^d$) is compared with ratios of the structure function of each nucleus to that of the free proton and free neutron, separately. The $F_2^A/F_2^d$ ratios for the carbon and gold nuclei are shown in Fig.~\ref{fig:data_np_ratio} using the collective world data as published in Ref.~\cite{Malace}. The gold nucleus (heavier than carbon) exhibits a steeper slope, a standard observation of the EMC Effect. The $F_2^A/AF_2^p$ ratio is the per nucleon ratio to the proton structure function and is also shown in Fig.~\ref{fig:data_np_ratio}(a). These ratios are consistent with theory predictions and exhibit a similar trend to the ratio of the nuclear structure functions to deuterium in that the slope steepens for heavier nuclei. The vertical spread in the ratios of the various nuclei softens when an iso-scalar correction is applied. The $F_2^p$ structure function is obtained from the CJ15 fit. The $F_2^A/AF_2^n$, shown in Fig.~\ref{fig:data_np_ratio}(a), is the per nucleon ratio with respect to the free neutron. Here, we observe a positive slope such that the heavier nuclei exhibit shallower slopes than the lighter nuclei.   
 
The $F_2^A/AF_2^n$ ratio is the per nucleon to the free neutron. The free neutron $F_2^n$ is from the CJ15 extraction from the world data. As shown in Fig.~\ref{fig:data_np_ratio}, the $F_2^A/AF_2^n$ ratios for the SLAC E139 nuclei exhibit a larger vertical spread when compared to the vertical spread of the $F_2^A/AF_2^p$ ratios for various nuclei. The $F_2^A/AF_2^n$ ratios also exhibit a positive slope in contrast to the ratios constructed using the free proton. These observations are further magnified in Fig.~\ref{fig:data_np_ratio}(b) where the ratios for each nuclei are shown as per proton ($F_2^A/ZF_2^p$) and per neutron ($F_2^A/NF_2^n$). The per proton ratios taken with respect to the free proton yield ratios that are nearly indistinguishable when compared to the large spread of the per neutron ratios taken with respect to the free neutron. Again, the per neutron slopes in Fig.~\ref{fig:data_np_ratio}(b) indicate a shallower slope (always positive) for the heavier nuclei.  
 
\section{Deuterium nuclear effects and kinematic dependence}
\subsection{Deuterium}
\begin{figure}[H]
\begin{minipage}{0.46\textwidth}
\setlabel{pos=see,fontsize=\scriptsize,labelbox}
 \xincludegraphics[width=\textwidth,label=(a),labelbox=false,fontsize=\Large]{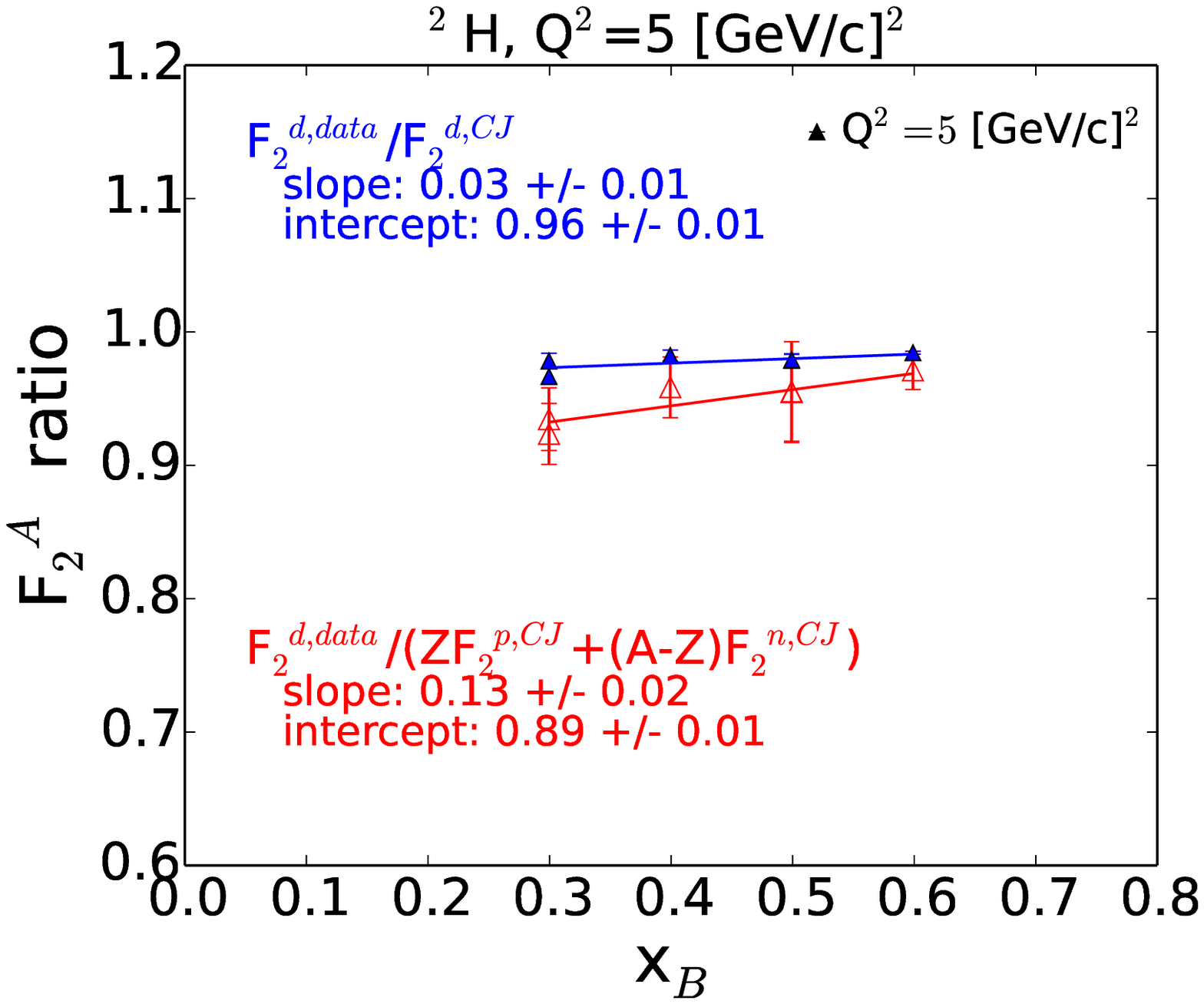}
\end{minipage}\hfill\begin{minipage}{0.46\textwidth}
\setlabel{pos=see,fontsize=\scriptsize,labelbox}
\xincludegraphics[width=\textwidth,label=(b),labelbox=false,fontsize=\Large]{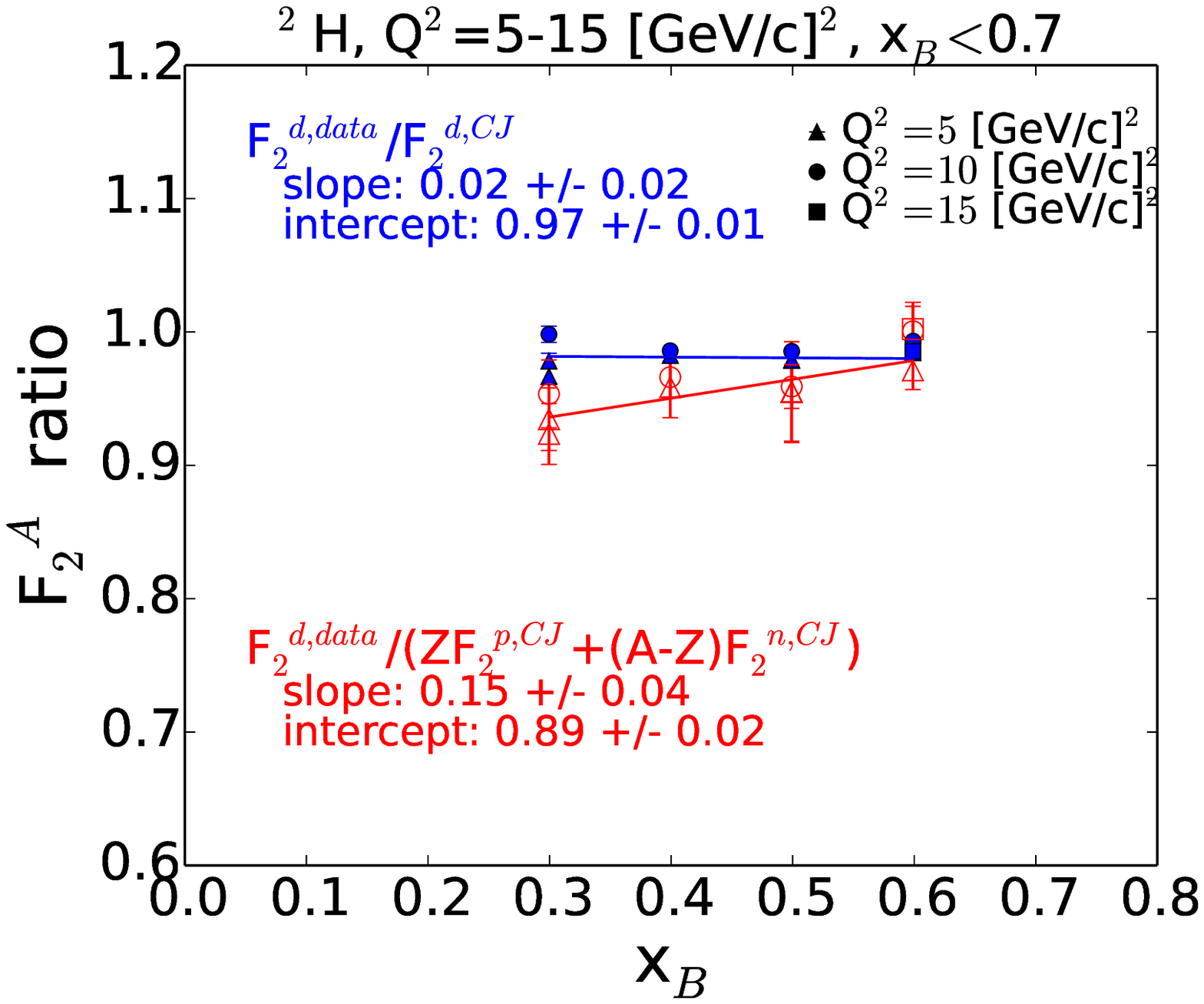}
\end{minipage}\hfill\begin{minipage}{0.46\textwidth}
\setlabel{pos=see,fontsize=\scriptsize,labelbox}
\xincludegraphics[width=\textwidth,label=(c),labelbox=false,fontsize=\Large]{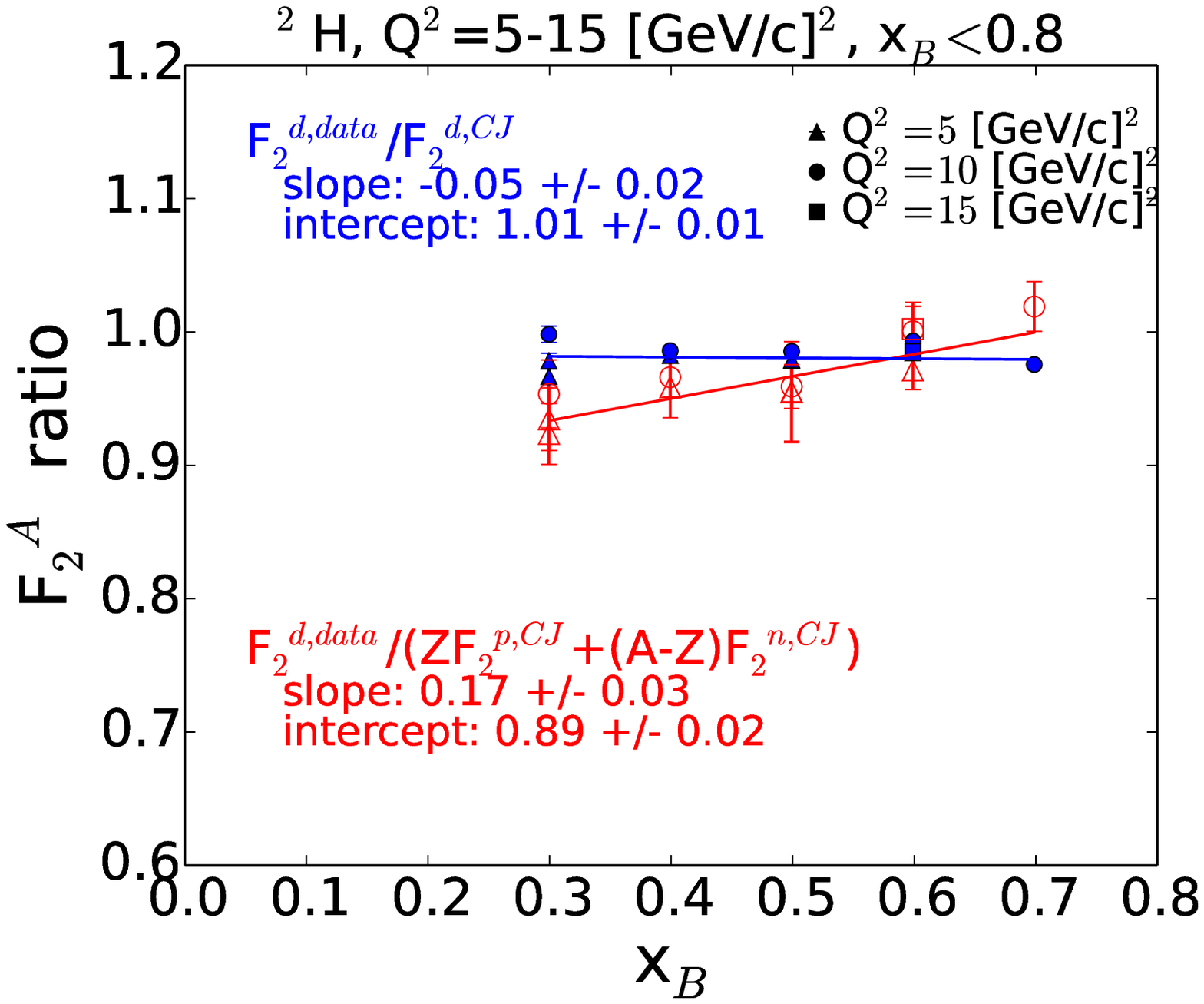}
\end{minipage}
  \caption[]{Linear fits to the deuterium target data with cuts on $Q^2$ and $x_B$. The blue solid points show the ratio of the E139 deuterium data relative to the CJ15 deuterium while the red open points show the ratio of the E139 deuterium data relative to the sum of the free proton and free neutron (excludes nuclear effects of deuterium). (a) Data taken for $Q^2=5$~GeV$^2$/c$^2$. (b) Data taken for all $Q^2$ but excluding the data at $x_B=0.7$. (c) All data taken through $x_B=0.7$.}
  \label{fig:fits_D}
\end{figure}

The nuclear effects in deuterium are non-negligible for values of $x_B$ from 0.3--0.7 in the region of interest to the EMC Effect. The nuclear effects in deuterium can be quantified somewhat as the difference between the sum of the free proton and free neutron to that of the measured deuterium structure function. In Fig.~\ref{fig:fits_D}, the $F_2^A$ ratio is defined in two ways: the standard $F_2^A/F_2^d$ (shown in blue, solid points) and the ratio to the sum of the free neutron and free proton $F_2^A/(ZF_2^p+(A-Z)F_2^n)$ (shown in red, open points). The SLAC E139 data was taken for $Q^2$ of 5, 10, and 15~GeV$^2$/c$^2$.

In Fig.~\ref{fig:fits_D}, the deuterium to nucleon ratios from the SLAC E139 deuterium data are shown using the CJ15 structure function constructs in the $F_2$ nucleon denominators. The different $Q^2$ data are shown using different shapes in each plot, and fits to the ratios are taken separately with and without the highest $x_B$ point at 0.7. There is a small systematic offset from unity for deuterium taken as a ratio to itself, but the ratio is relatively flat across $x_B$. There exists a clear difference between the ratio taken relative to the deuteron structure function (blue, solid) and the ratio taken relative to sum of the free neutron and free proton structure functions (red, open), a measure of the nuclear effects in the deuteron. This difference also exhibits a small spread in $Q^2$. We note the important changes when different $x_B$, $Q^2$ cuts are made to the data; reflecting the kinematic dependence in both $x_B$ and $Q^2$ of deuteron nuclear effects. 

\subsection{Heavier nuclei}

\begin{figure}[H]
\begin{center}
\includegraphics[width=0.46\textwidth]{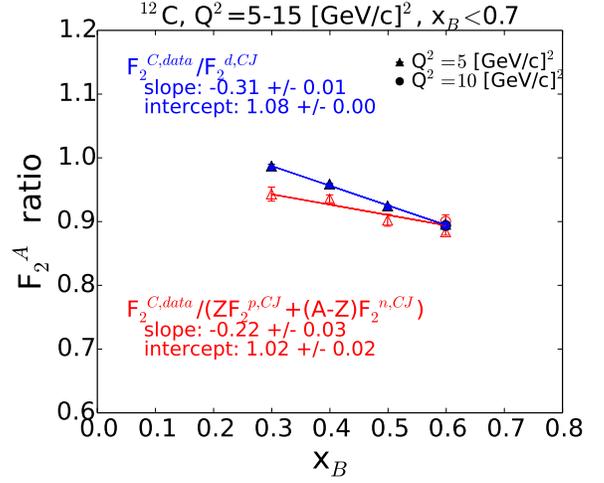}
  \caption[]{Linear fits to the $C$ target data with cuts on $Q^2$ and $x_B$. The blue solid data points show the ratio of the nucleus relative to deuterium while the red open data points show the ratio of the nucleus relative to the sum of the free proton and free neutron (excluding the nuclear effects of deuterium). The linear fit to the data is taken for all $Q^2$, where $5<Q^2<15$~GeV/c$^2$. No data was taken in the E139 experiment for carbon at $x_B=0.7$.}
  \label{fig:fits_C}
  \end{center}
\end{figure}

\begin{figure}[H]
\begin{minipage}{0.46\textwidth}
\setlabel{pos=see,fontsize=\scriptsize,labelbox}
 \xincludegraphics[width=\textwidth,label=(a),labelbox=false,fontsize=\Large]{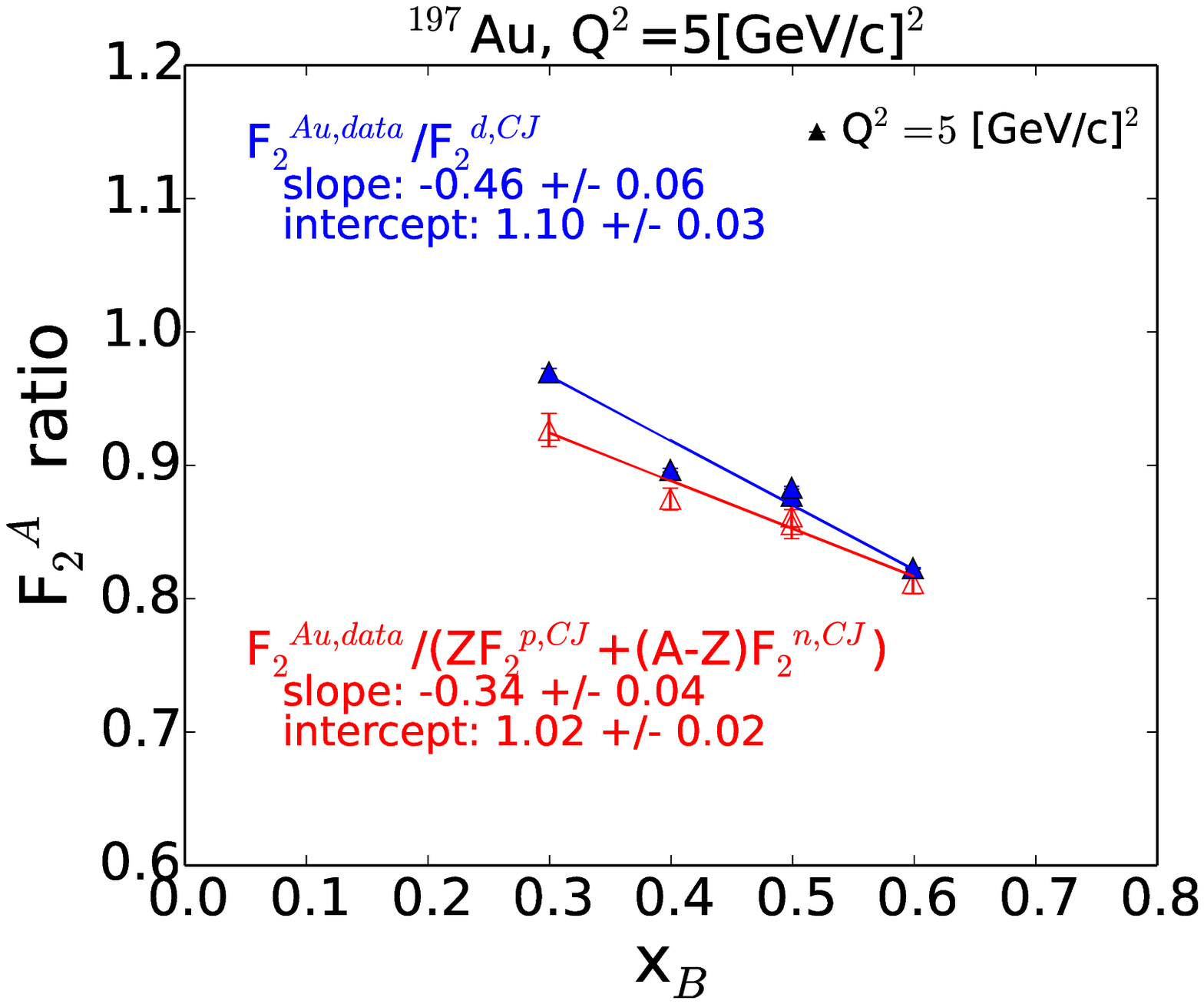}
\end{minipage}\hfill\begin{minipage}{0.46\textwidth}
\setlabel{pos=see,fontsize=\scriptsize,labelbox}
\xincludegraphics[width=\textwidth,label=(b),labelbox=false,fontsize=\Large]{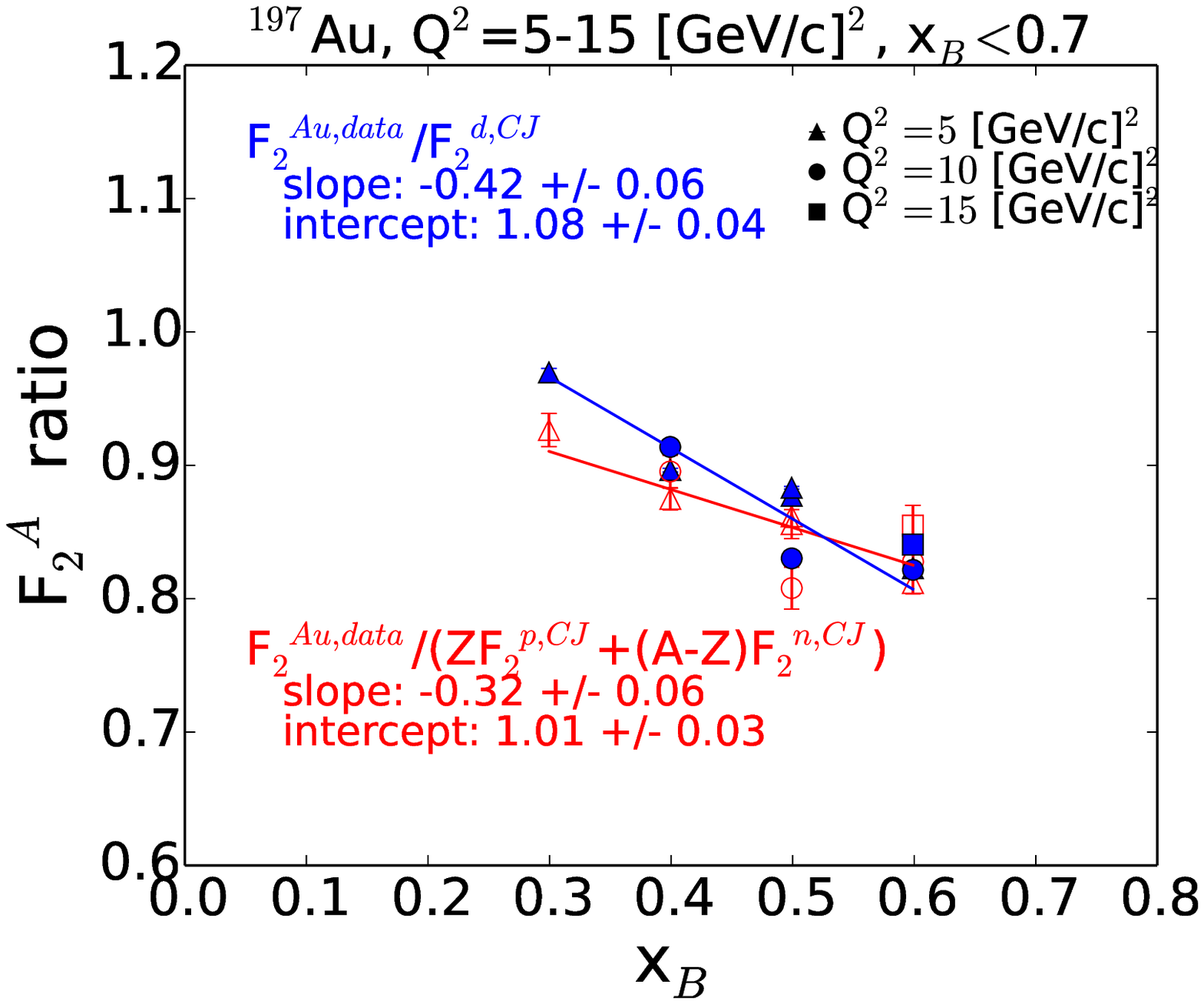}
\end{minipage}\hfill\begin{minipage}{0.46\textwidth}
\setlabel{pos=see,fontsize=\scriptsize,labelbox}
\xincludegraphics[width=\textwidth,label=(c),labelbox=false,fontsize=\Large]{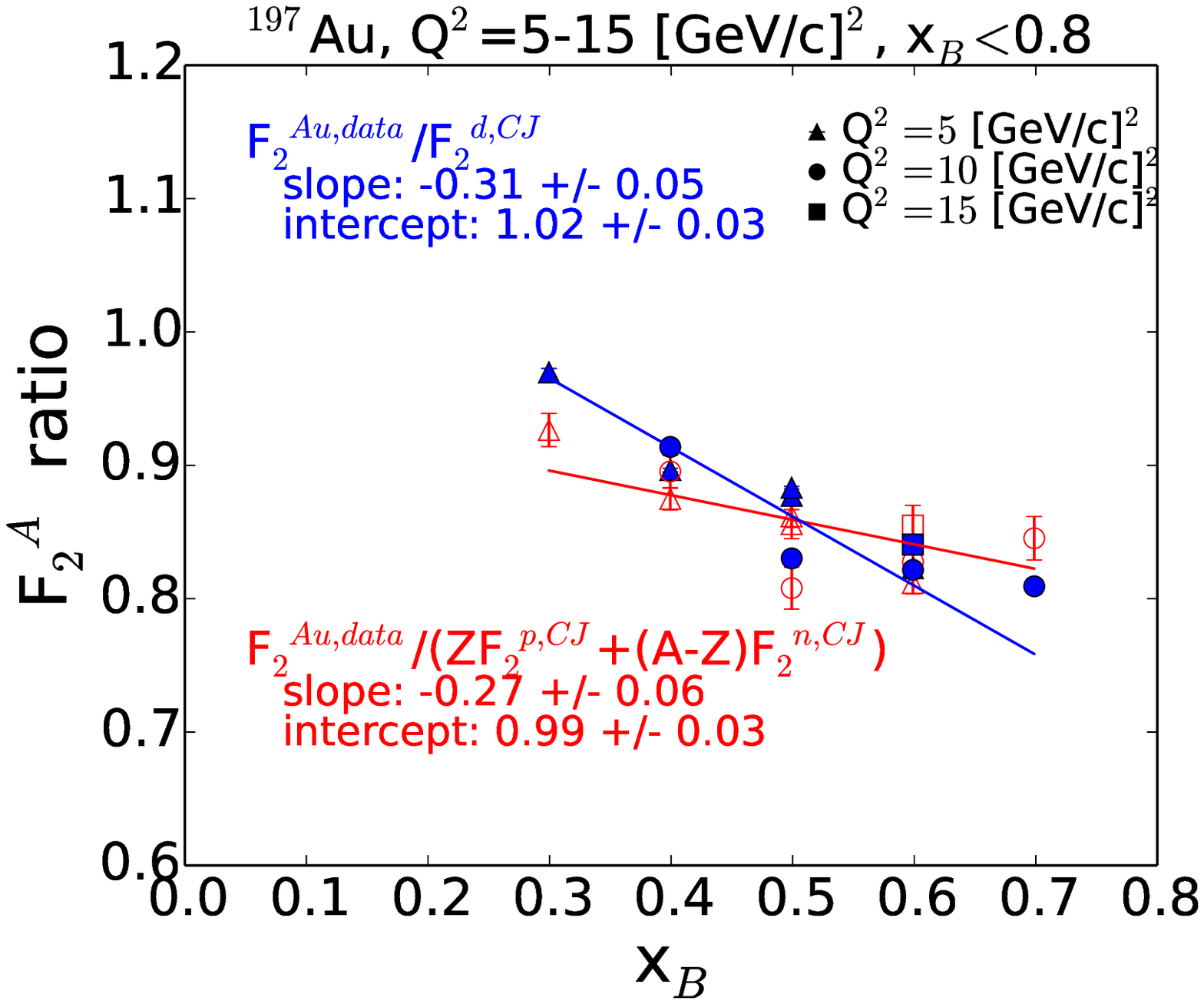}
\end{minipage}
  \caption[]{Linear fits to the $Au$ target data with cuts on $Q^2$ and $x_B$. The blue solid data points show the ratio of the nucleus relative to deuterium while the red data points show the ratio of the nucleus relative to the sum of the free proton and free neutron (excluding nuclear effects of deuterium). (a) linear fit to the data taken for $Q^2=5$~GeV$^2$/c$^2$. (b) linear fit to the data taken for all $Q^2$ but excluding the data at $x_B=0.7$. (c) linear fit to all data taken through $x_B=0.7$.}
  \label{fig:fits_Au}
\end{figure}

The EMC ratios for all the target nuclei in the SLAC E139 experiment are compared for ratios where the denominator is deuterium and where the denominator is the sum of the free proton and neutron contributions. This comparison provides a measure of the contribution of deuterium nuclear corrections to the EMC Effect. Carbon is a symmetric nucleus, and the results are shown in Fig.~\ref{fig:fits_C}. Also shown in Fig.~\ref{fig:fits_C} is a linear fit to the two types of ratios, over the EMC range of data at $x_B=0.3$ and above.
\noindent No data was taken in the E139 experiment for carbon at $x_B>0.6$, where deuteron nuclear effects increase. The slope is more shallow for the ratio taken with the free neutron and proton components only. A more extreme case is seen in the heavy and asymmetric gold nucleus in Fig.~\ref{fig:fits_Au} in Section~\ref{appendix}. 

For the gold nucleus, the slope of the fit to ratios obtained with respect to the free neutron and free proton is shallower than the slope with respect to deuterium (as consistent with the general trend observed in carbon and deuterium as previously discussed). When data is fitted for a range of $Q^2$ values, a noticeable striping at each point in $x_B$ is apparent indicating some $Q^2$ dependence. Furthermore, the higher $x_B$ data (also higher $Q^2$ due to the experimental kinematics in scattering angle), is barely linear with respect to the lower $x_B$ data. 

\begin{figure}[H]
\begin{minipage}{0.46\textwidth}
\setlabel{pos=sww,fontsize=\scriptsize,labelbox}
 \xincludegraphics[width=\textwidth,label=(a),labelbox=false,fontsize=\Large]{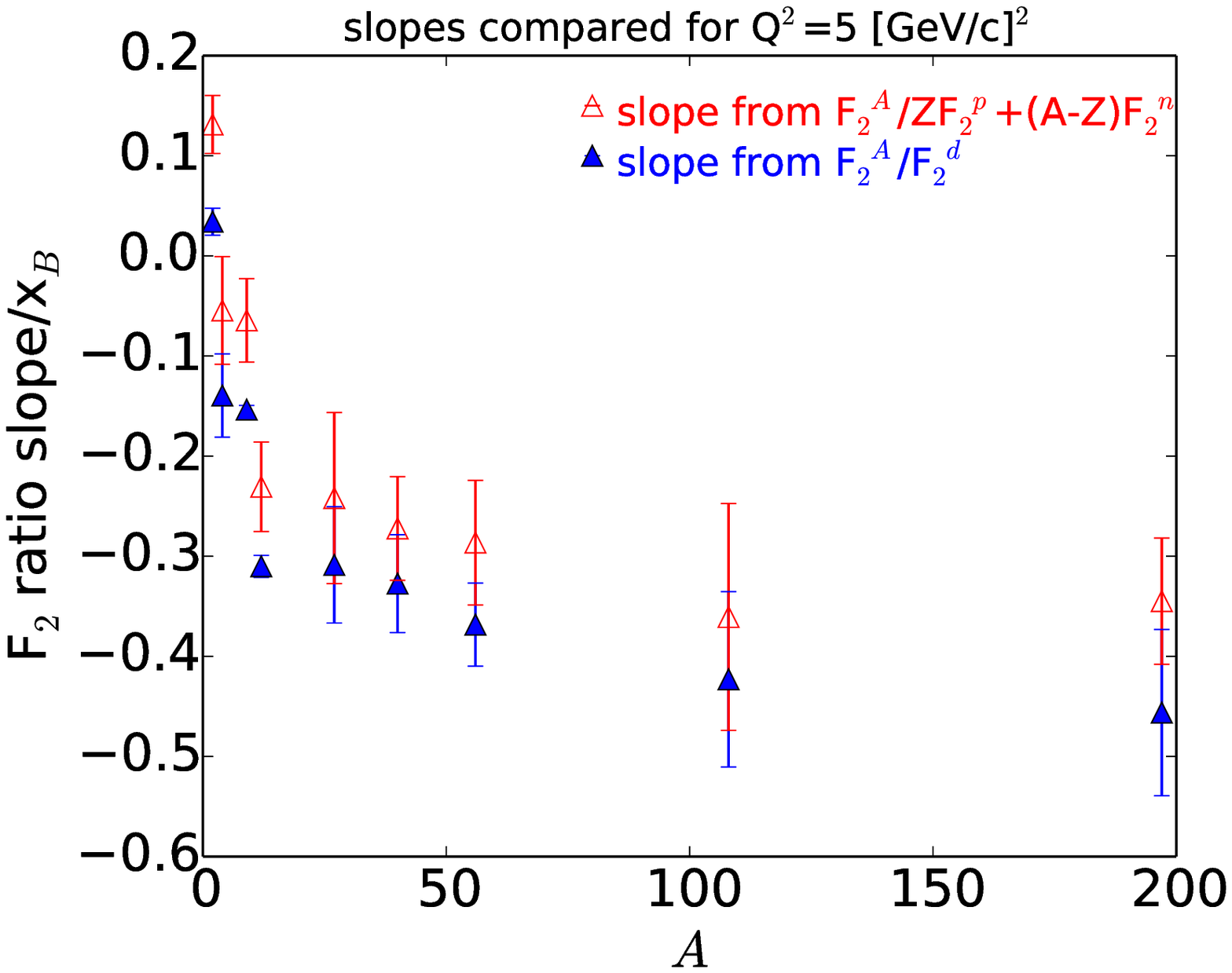}
\end{minipage}\hfill\begin{minipage}{0.46\textwidth}
\setlabel{pos=sww,fontsize=\scriptsize,labelbox}
\xincludegraphics[width=\textwidth,label=(b),labelbox=false,fontsize=\Large]{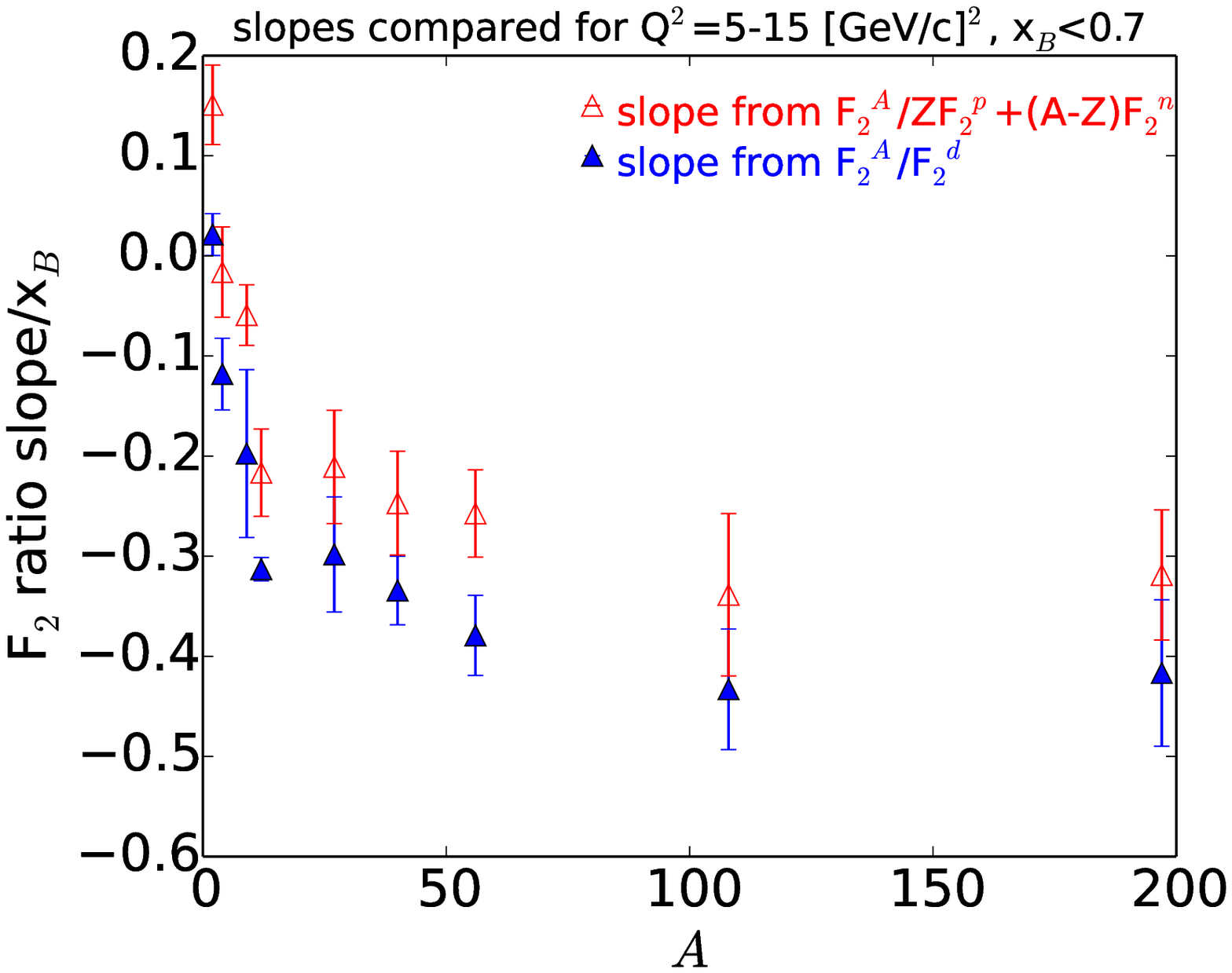}
\end{minipage}\hfill\begin{minipage}{0.46\textwidth}
\setlabel{pos=sww,fontsize=\scriptsize,labelbox}
\xincludegraphics[width=\textwidth,label=(c),labelbox=false,fontsize=\Large]{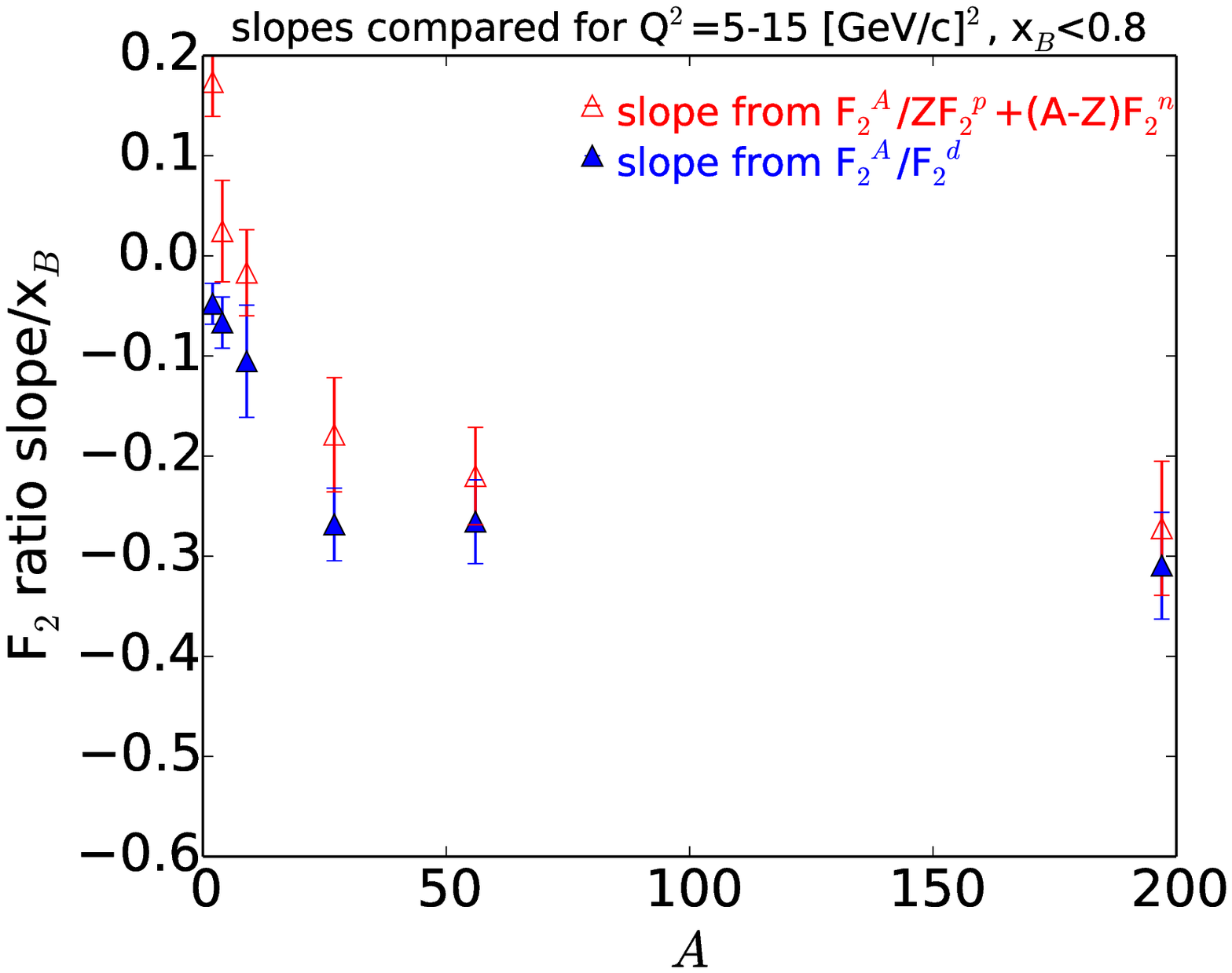}
\end{minipage}
  \caption[]{The slopes to the fits of the $F_2$ ratios versus $x_B$ are shown as a function of the target $A$. (a) The slope comparisons are made using all data with $Q^2=5$~[GeV/c]$^2$. (b) The slopes are compared with no cut on $Q^2$ but requiring that $x_B<0.7$. (c) The slopes are compared with no cut on $Q^2$ and $x_B<0.8$.}
  \label{fig:Aslope_summary}
\end{figure} 
\noindent In all nuclei (see Section~\ref{appendix}), this discrepancy is most apparent for increasing values of $x_B$ around 0.7. This observation is where one would expect to see the contributions from the deuteron nuclear effects as shown in Fig.~\ref{fig:dn_theory}. 

Similar fits are made to all the nuclei (shown in Section~\ref{appendix}). The resultant summary of the slopes is shown in Tables~\ref{SlopeFits}, \ref{SlopeFits1}, and \ref{SlopeFits2} in Section~\ref{appendix}. The slopes are shown as a function of $A$ for the nuclei in  Fig.~\ref{fig:Aslope_summary}.

\begin{figure}[H]
\begin{minipage}{0.46\textwidth}
\setlabel{pos=sww,fontsize=\scriptsize,labelbox}
 \xincludegraphics[width=\textwidth,label=(a),labelbox=false,fontsize=\Large]{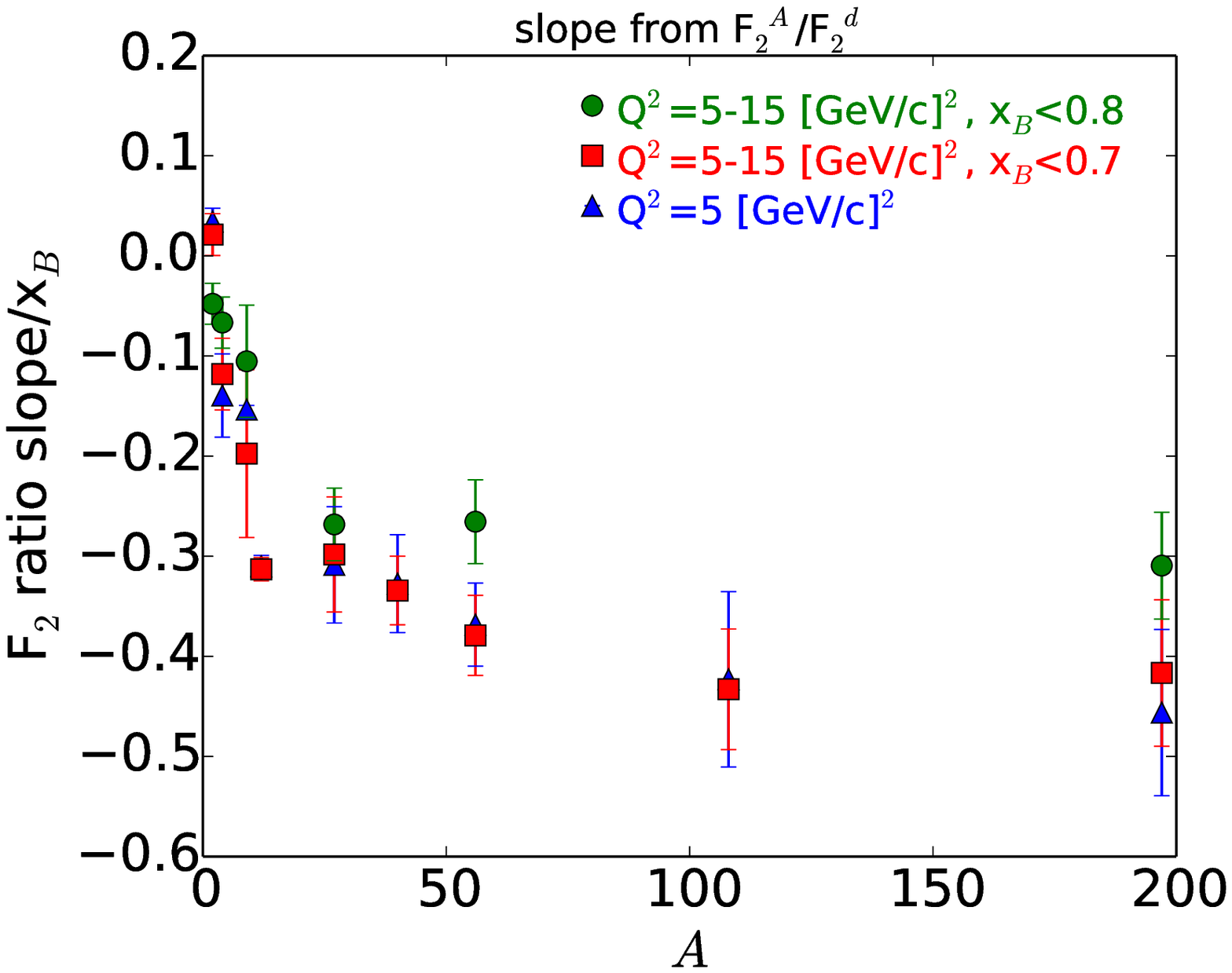}
\end{minipage}\hfill\begin{minipage}{0.46\textwidth}
\setlabel{pos=sww,fontsize=\scriptsize,labelbox}
\xincludegraphics[width=\textwidth,label=(b),labelbox=false,fontsize=\Large]{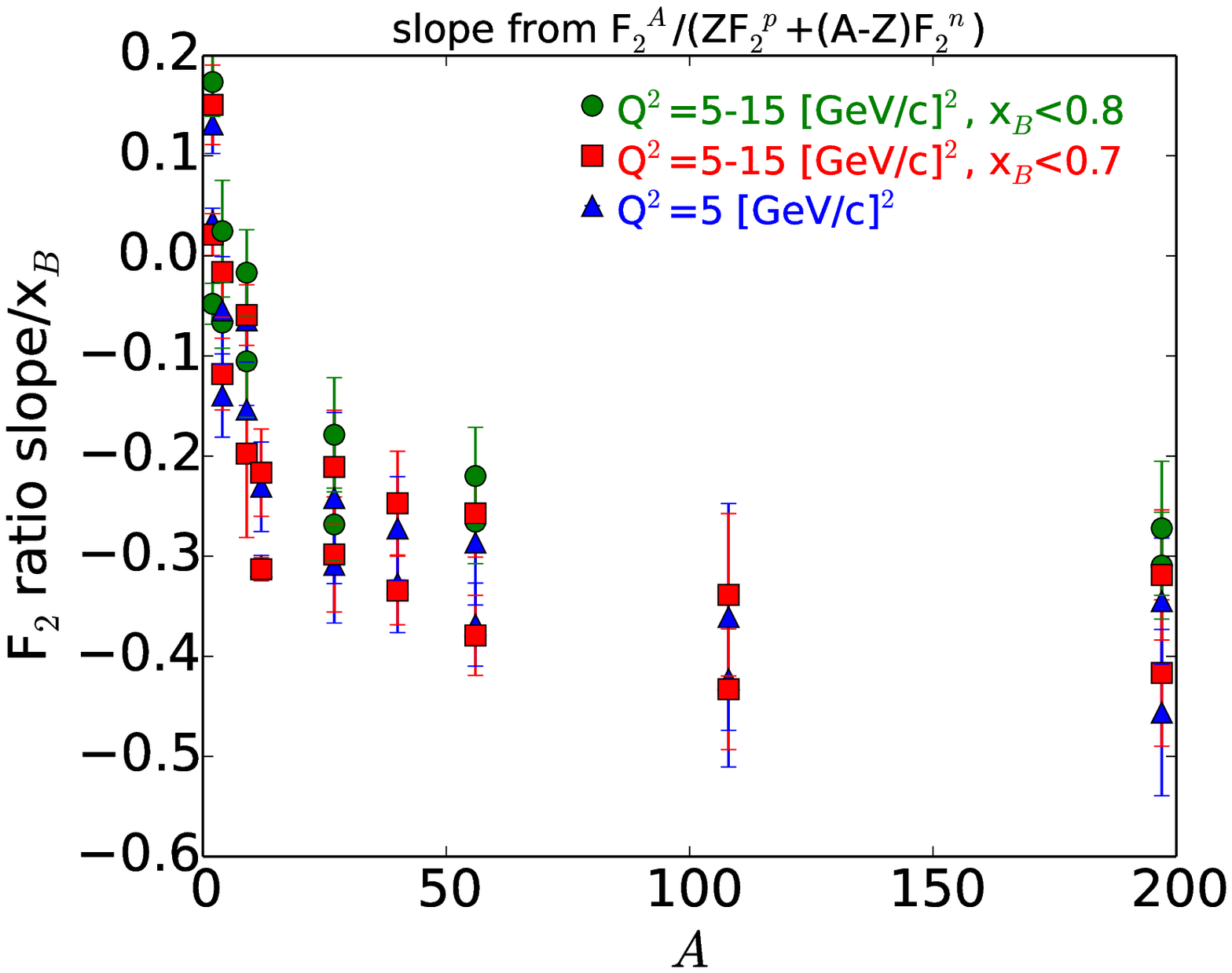}
\end{minipage}
  \caption[]{The slopes are compared for the different cuts in $Q^2$ and $x_B$. The inclusion of larger $Q^2$ and $x_B$ points in the fit generates a somewhat shallower slope.}
  \label{fig:Aslope_compare}
\end{figure} 

The slopes for each ratio for various cuts in $Q^2$ and $x_B$ are compared in Fig.~\ref{fig:Aslope_compare}. The inclusion of higher $Q^2$ and $x_B$ data in the fits generates somewhat shallower slopes due to the rise seen from the nuclear effects. The inclusion of the largest $x_B$, $Q^2$ yields a deuterium $F_2^A/F_2^d$ratio closest to one even though this is the region of largest deuteron nuclear correction. In all cases, the ratios to the free neutron and proton (b) have a reduced EMC Effect (i.e. shallower slopes) than the deuteron ratios(a). Linear fits are sufficient to describe the data. Including the largest $x_B$ data further reduces the slopes, underscoring the large effect of deuteron nuclear corrections, which increase with increasing $x_B$. 

Table~\ref{SlopeFits} (see Section~\ref{appendix}) shows the fitted slope and intercept values for both the $F_2^A$ ratio for each nucleus taken with respect to deuterium and the sum of the free neutron and free proton contributions. In Table~\ref{SlopeFits}, only fits to the data where $Q^2=5$~GeV$^2$/c$^2$ are shown. No iso-scalar corrections are applied to the asymmetric nuclei.

\begin{figure}[H]
  \centering
      	  \includegraphics[width=0.48\textwidth]{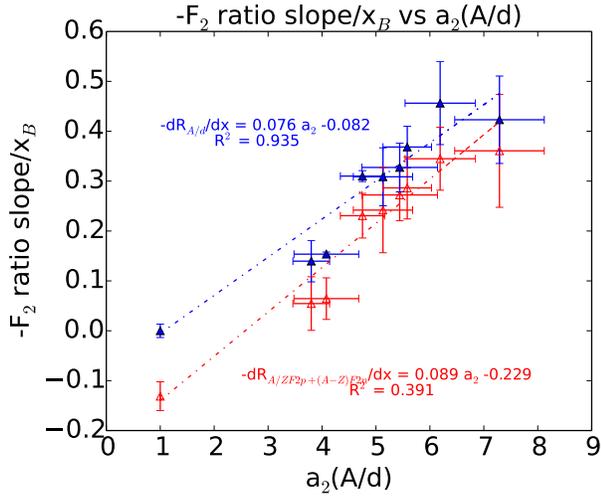}
 	 \caption[]{The slopes of the fit to the data for $Q^2=5$~GeV$^2$/c$^2$ are shown versus the $a_2$ SRC factor from~\cite{Weinstein}. The points shown in blue (solid) are the slopes from the ratios taken with respect to deuterium, and the points in red (open) are from the slopes with respect to the free neutron and free proton denominator. The slope value for deuterium taken with respect to itself (blue) is set to 0.}
  \label{fig:a2_src}
 \end{figure}

Table~\ref{SlopeFits1} shows the fitted slope and intercept values for both the $F_2^A$ ratios where all E139 $Q^2$ data are included at $x_B<0.7.$ Table~\ref{SlopeFits2} shows the fitted slope and intercept values for both the $F_2^A$ ratios where all E139 $Q^2$ data are included with no cuts on $x_B<0.8$. In all three tables, the fits were taken excluding data below $x_B$ of 0.3. 

 The slopes from Table~\ref{SlopeFits} where $Q^2=5$~GeV$^2$/c$^2$ only are shown plotted against the short range correlated (SRC) scale factor $a_2$~\cite{Weinstein}, derived as the ratio $A/d$ in the plateau region of large $x_B$ and can be summarized as the probability that a nucleon belongs to an SRC pair. The general fitted slope is still linear across the $a_2$ values for each nucleus, but there is a clear systematic offset between the intercept values of these two fits. The fits to the slopes with respect to the SRC factor are shown in Fig.~\ref{fig:a2_src}. It should be noted that the $a_2$ values shown here include predicted values from~\cite{Weinstein}. While some work has been done to improve these values from later analysis, \cite{Weinstein} provides the most complete list for the nuclei of interest and was used for consistency.

 
The intercept value of approximately -0.08 for the ratio taken with respect to deuterium (Fig.~\ref{fig:a2_src}) is consistent with previous studies~\cite{Weinstein}~\cite{Hen}. The intercept value is significantly smaller for the data where the slope ratio was taken with respect to the sum of the free proton and free neutron contributions and thus yields a smaller intercept value of approximately -0.23. While later analysis has yielded some slightly different calculations for the $a_2$ values, the overall differences between the slopes and intercept will remain consistent.

\section{Conclusions}

The CJ Collaboration has recently extracted the free neutron structure function from the world DIS data, thereby opening new windows for exploring the EMC Effect and deuteron nuclear effects. General observations show that the deuteron ratio to the sum of the free proton and free neutron structure functions has some dependence on $Q^2$, most significantly at large $Q^2$ and large $x_B$. As the EMC Effect is often obtained from the slope of the nuclear to deuteron structure function ratios over a range in large $x_B$, there are kinematically-dependent nuclear effects in this regime from the deuteron alone. Such dependence may have been minimized in previous studies as the nuclear data were averaged in $Q^2$. Additionally, the correlation of these effects in $x_B$ and $Q^2$ dictate that the structure function ratios obtained using data at $x_B>0.6$ do not follow the same linear trend of the data from lower $x_B$. The EMC Effect as calculated using the slope of the $F_2^A/F_2^d$ ratio has an $x_B$ and $Q^2$ dependence. Also, there appears to be a more significant spread in $A$ in the data where the structure function of the nucleus is taken with respect to the neutron as compared to the proton. This spread arises from the drop in the ratio $F_2^n/F_2^p$ as a function of $x_B$. Moreover, by removing the deuteron nuclear effects, a correlation is still observed between the resultant EMC slope and short range correlated pairs in the nucleus, but the overall slope of this correlation is modified slightly, yielding a smaller intercept. The authors of this paper would be interested to further investigate avenues for disentangling the $x_B$ and $Q^2$ data at higher $x_B$ to better understand the nuclear effects in this regime. 

\section{Acknowledgments}

The authors would like to thank the CJ Collaboration and especially Shujie Li for constructing the most complete world $F_2^n$ data set to date. The authors also with to thank Ian Cloet, Alberto Accardi, and Wally Melnitchouk for their theory support and providing crucial insights for this analysis.  This work was supported in part by research grant 1700333 from the National Science Foundation. This work is supported by the U.S. Department of Energy, Office of Science, Office of Nuclear Physics under contract DE-AC05-06OR23177.

\bibliography{EMCnote}{}
\bibliographystyle{unsrt}


\newpage
\section{Appendix}\label{appendix}

\begin{table}[H]
\caption{\label{SlopeFits} Summary of linear fits to ratios as a function of $x_B$ where $Q^2=5$~GeV$^2$/c$^2$.}
\centering
\begin{tabular}{ | C{1cm} | C{2.3cm} | C{2.3cm} | C{2.3cm} | C{2.3cm} | }
 \hline
 \textbf{Nuc.} & \textbf{A/d slope} & \textbf{A/d intercept} & \textbf{A/(n+p) slope} & \textbf{A/(n+p) intercept} \\ 
  \hline
$^2$H & 0.03+/-0.01	& 0.96+/-0.01 & 0.13+/-0.02 &	0.89+/-0.01\\ 
  \hline
  $^4$He & -0.14+/-0.03	 & 1.01+/-0.02 & -0.05+/-0.04 & 0.95+/-0.02 \\ 
 \hline
 $^9$Be & -0.15+/-0.00	 &1.02+/-0.00 &-0.06+/-0.02&	0.96+/-0.01\\ 
  \hline
   $^{12}$C & -0.31+/-0.01	& 1.08+/-0.00 & -0.22+/-0.03	&1.02+/-0.02  \\ 
  \hline
    $^{27}$Al &-0.31+/-0.03 &	1.07+/-0.02 & -0.24+/-0.05 &	1.02+/-0.02 \\ 
  \hline
 $^{40}$Ca &-0.33+/-0.03 &	1.09+/-0.02 & -0.27+/-0.03 &	1.04+/-0.01 \\ 
  \hline  
  $^{56}$Fe & -0.37+/-0.03&	1.09+/-0.02& -0.29+/-0.05 &	1.02+/-0.02 \\ 
  \hline 
  $^{108}$Ag & -0.42+/-0.05 &	1.12+/-0.03 & -0.36+/-0.07	& 1.07+/-0.03 \\ 
  \hline 
   $^{197}$Au & -0.46+/-0.06	&1.10+/-0.03 & -0.34+/-0.04 &	1.02+/-0.02\\ 
  \hline 
    \end{tabular}
\end{table}

\begin{table}[H]
\caption{\label{SlopeFits1} Summary of linear fits to ratios as a function of $x_B$ where $Q^2=5-15$~GeV$^2$/c$^2$ and $x_B<0.7$.}
\centering
\begin{tabular}{ | C{1cm} | C{2.3cm} | C{2.3cm} | C{2.3cm} | C{2.3cm} | }
 \hline
 \textbf{Nuc.} & \textbf{A/d slope} & \textbf{A/d intercept} & \textbf{A/(n+p) slope} & \textbf{A/(n+p) intercept} \\ 
  \hline
$^2$H & 0.02+/-0.02 &	0.97+/-0.01 & 0.15+/-0.04& 0.89+/-0.02 \\ 
  \hline
  $^4$He &-0.12+/-0.03 &	1.0+/-0.02 & -0.02+/-0.04 &	0.93+/-0.02 \\ 
 \hline
 $^9$Be &-0.20+/-0.07 &	1.04+/-0.04 &-0.06+/-0.03 &	0.95+/-0.01\\ 
  \hline
   $^{12}$C -0.31+/-0.01	& 1.08+/-0.00 & -0.22+/-0.03	&1.02+/-0.02 \\ 
  \hline
    $^{27}$Al &-0.30+/-0.05	&1.07+/-0.03 & -0.21+/-0.05	&1.01+/-0.02 \\ 
  \hline
 $^{40}$Ca &-0.33+/-0.02&	1.09+/-0.01 & -0.25+/-0.04 &	1.03+/-0.02 \\ 
  \hline  
  $^{56}$Fe & -0.38+/-0.04	&1.09+/-0.02& -0.26+/-0.04 &	1.01+/-0.02 \\ 
  \hline 
  $^{108}$Ag &-0.43+/-0.04&	1.12+/-0.02 & -0.34+/-0.06	&1.06/-0.03 \\ 
  \hline 
   $^{197}$Au & -0.42+/-0.06&	1.08+/-0.04 & -0.32+/-0.06&	1.01+/-0.03\\ 
  \hline 
    \end{tabular}
\end{table} 

\begin{table}[H]
\caption{\label{SlopeFits2} Summary of linear fits to ratios as a function of $x_B$ where $Q^2=5-15$ GeV$^2$/c$^2$ and $x_B<0.8$.}
\centering
\begin{tabular}{ | C{1cm} | C{2.3cm} | C{2.3cm} | C{2.3cm} | C{2.3cm} | }
 \hline
 \textbf{Nuc.} & \textbf{A/d slope} & \textbf{A/d intercept} & \textbf{A/(n+p) slope} & \textbf{A/(n+p) intercept} \\ 
  \hline
$^2$H & -0.05+/-0.02	&1.01+/-0.01& 0.17+/-0.03	&0.89+/-0.02 \\ 
  \hline
  $^4$He &-0.07+/-0.02&	0.97+/-0.01 & 0.02+/-0.04&	0.91+/-0.02 \\ 
 \hline
 $^9$Be &-0.11+/-0.05&	0.99+/-0.03 &-0.02+/-0.04&	0.93+/-0.02\\ 
  \hline
   $^{12}$C &---&--- & ---	&--- \\ 
  \hline
    $^{27}$Al &-0.27+/-0.03	&1.05+/-0.02&-0.18+/-0.05&	1.0+/-0.03 \\ 
  \hline
 $^{40}$Ca &---&--- & ---	&--- \\ 
  \hline  
  $^{56}$Fe &-0.27+/-0.04	&1.03+/-0.02&-0.22+/-0.04	&0.99+/-0.02\\ 
  \hline 
  $^{108}$Ag &---&--- & ---	&---  \\ 
  \hline 
   $^{197}$Au & -0.31+/-0.05 &	1.02+/-0.03 & -0.27+/-0.06 &	0.99+/-0.03\\ 
  \hline 
    \end{tabular}
\end{table}

\clearpage

\begin{figure}[H]
\begin{minipage}{0.46\textwidth}
\setlabel{pos=see,fontsize=\scriptsize,labelbox}
 \xincludegraphics[width=\textwidth,label=(a),labelbox=false,fontsize=\Large]{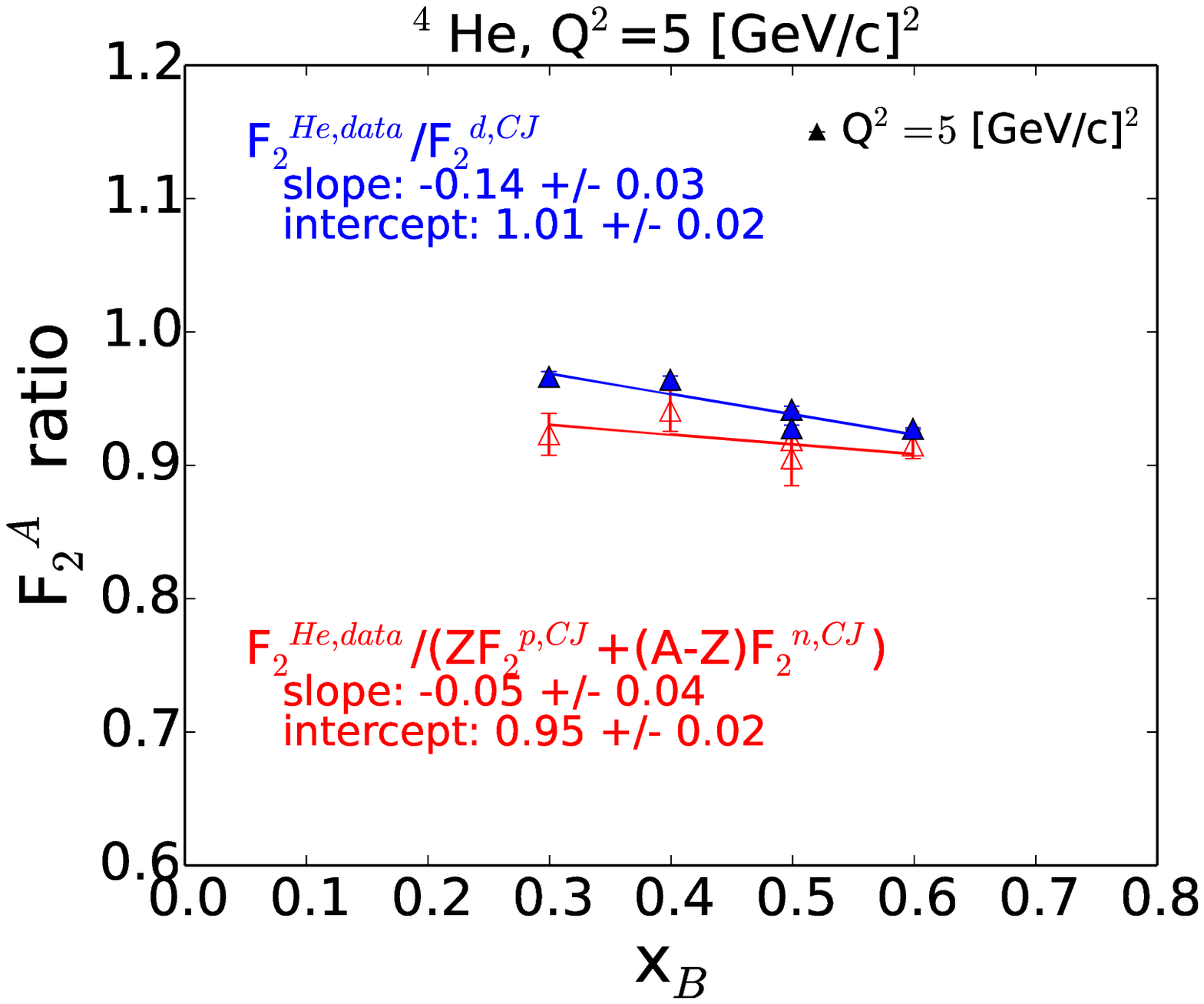}
\end{minipage}\hfill\begin{minipage}{0.46\textwidth}
\setlabel{pos=see,fontsize=\scriptsize,labelbox}
\xincludegraphics[width=\textwidth,label=(b),labelbox=false,fontsize=\Large]{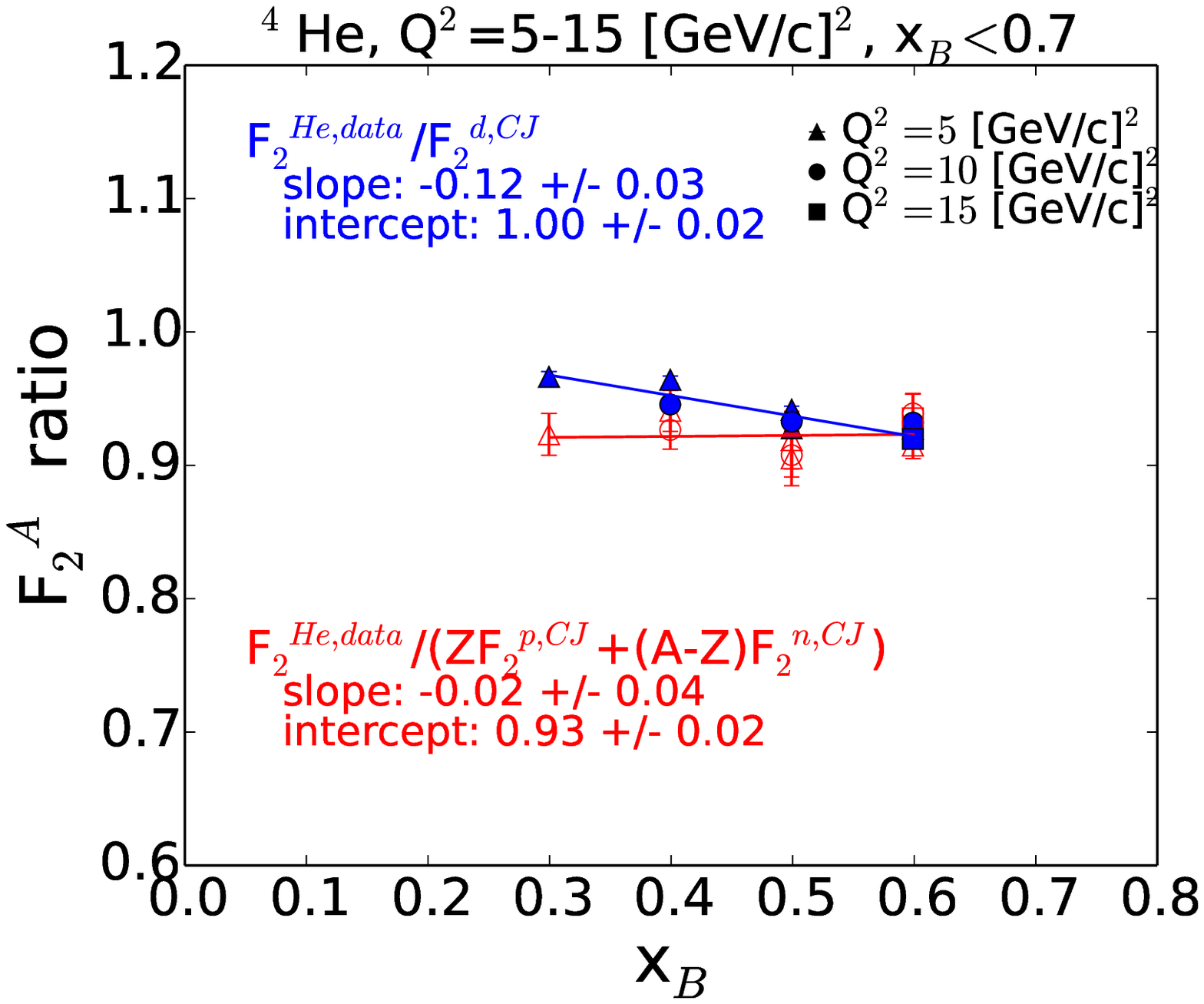}
\end{minipage}\hfill\begin{minipage}{0.46\textwidth}
\setlabel{pos=see,fontsize=\scriptsize,labelbox}
\xincludegraphics[width=\textwidth,label=(c),labelbox=false,fontsize=\Large]{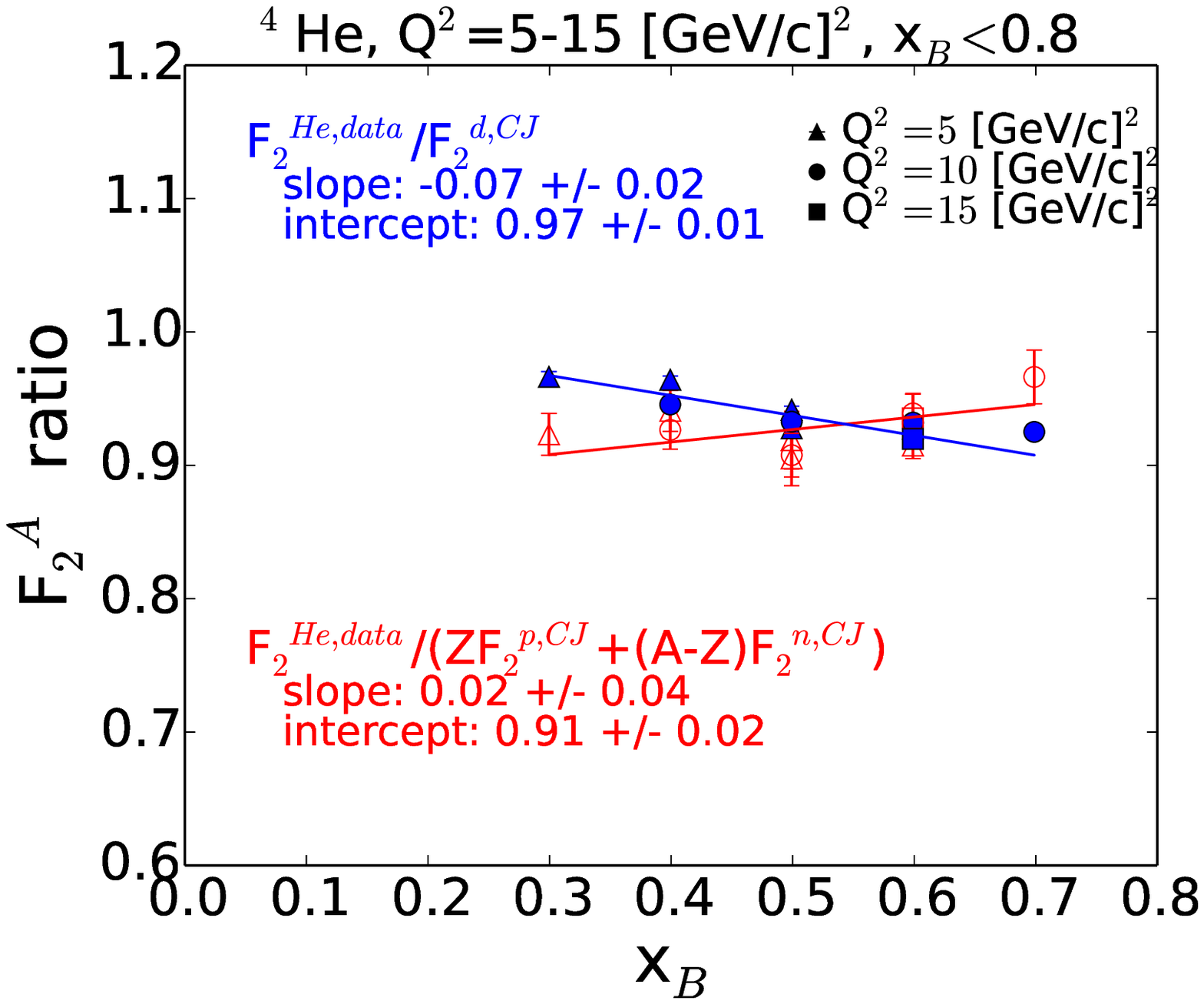}
\end{minipage}
  \caption[]{Linear fits to the $He$ target data with cuts on $Q^2$ and $x_B$ from the E139 experiment. (a) The ratios are constructed requiring that $Q^2=5$~GeV$^2$/c$^2$. (b) The ratios include all data $5<Q^2<15$~GeV$^2$/c$^2$ where $x_B<0.7$. (c) The ratios include all data $5<Q^2<15$~GeV$^2$/c$^2$ up to $x_B<0.8$.}
  \label{fig:fits_He}
\end{figure}   

 \begin{figure}[H]
\begin{minipage}{0.46\textwidth}
\setlabel{pos=see,fontsize=\scriptsize,labelbox}
 \xincludegraphics[width=\textwidth,label=(a),labelbox=false,fontsize=\Large]{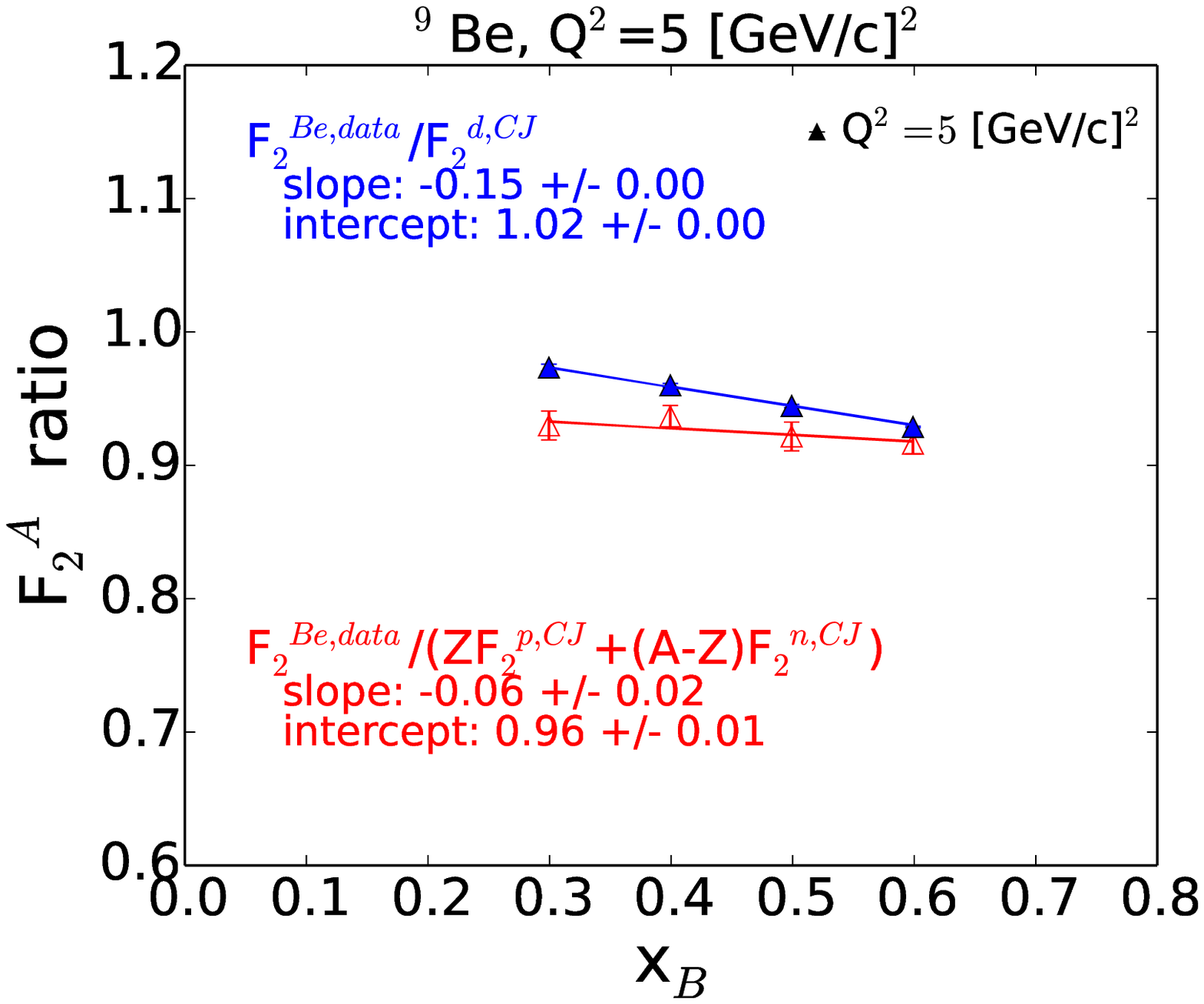}
\end{minipage}\hfill\begin{minipage}{0.46\textwidth}
\setlabel{pos=see,fontsize=\scriptsize,labelbox}
\xincludegraphics[width=\textwidth,label=(b),labelbox=false,fontsize=\Large]{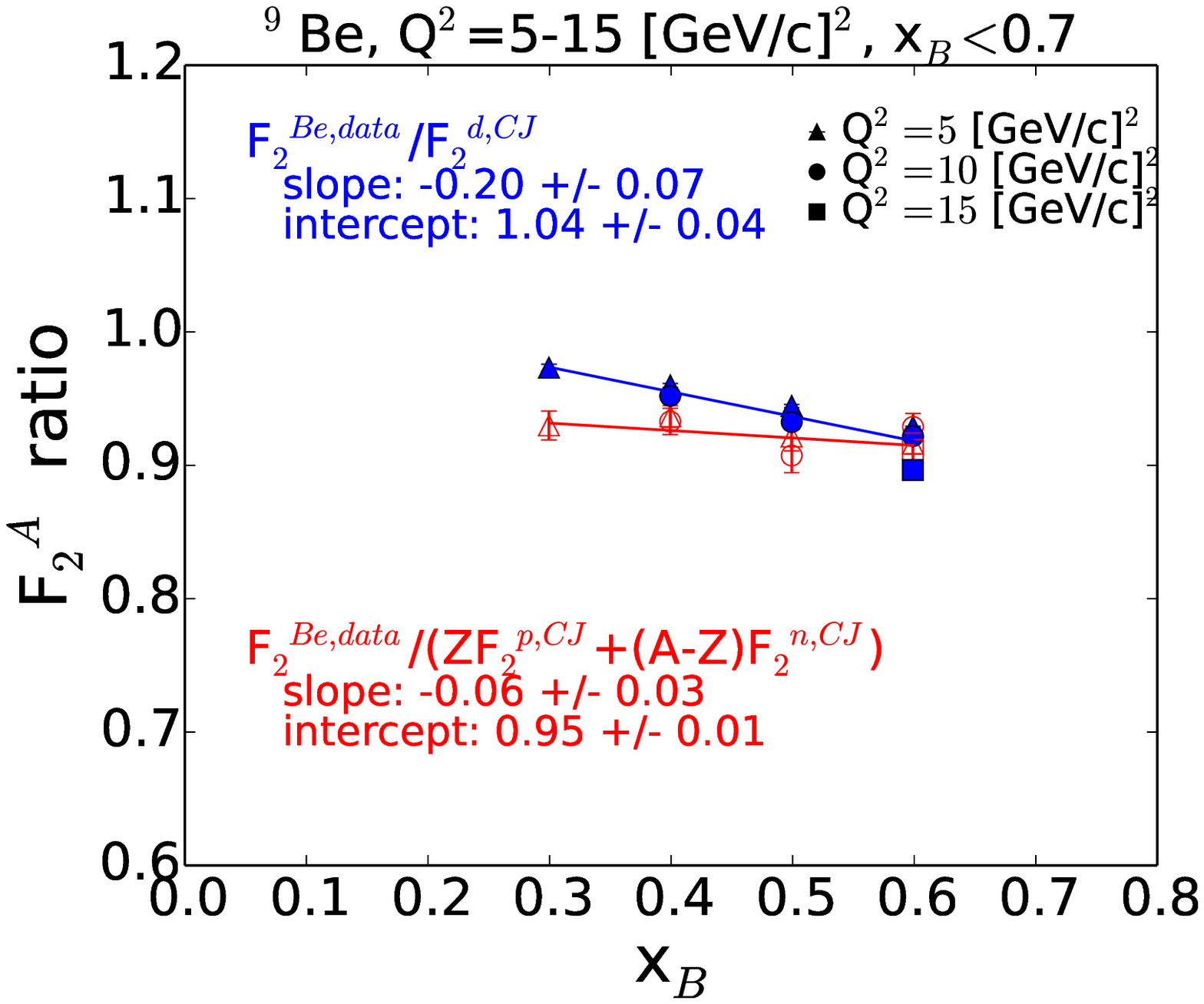}
\end{minipage}\hfill\begin{minipage}{0.46\textwidth}
\setlabel{pos=see,fontsize=\scriptsize,labelbox}
\xincludegraphics[width=\textwidth,label=(c),labelbox=false,fontsize=\Large]{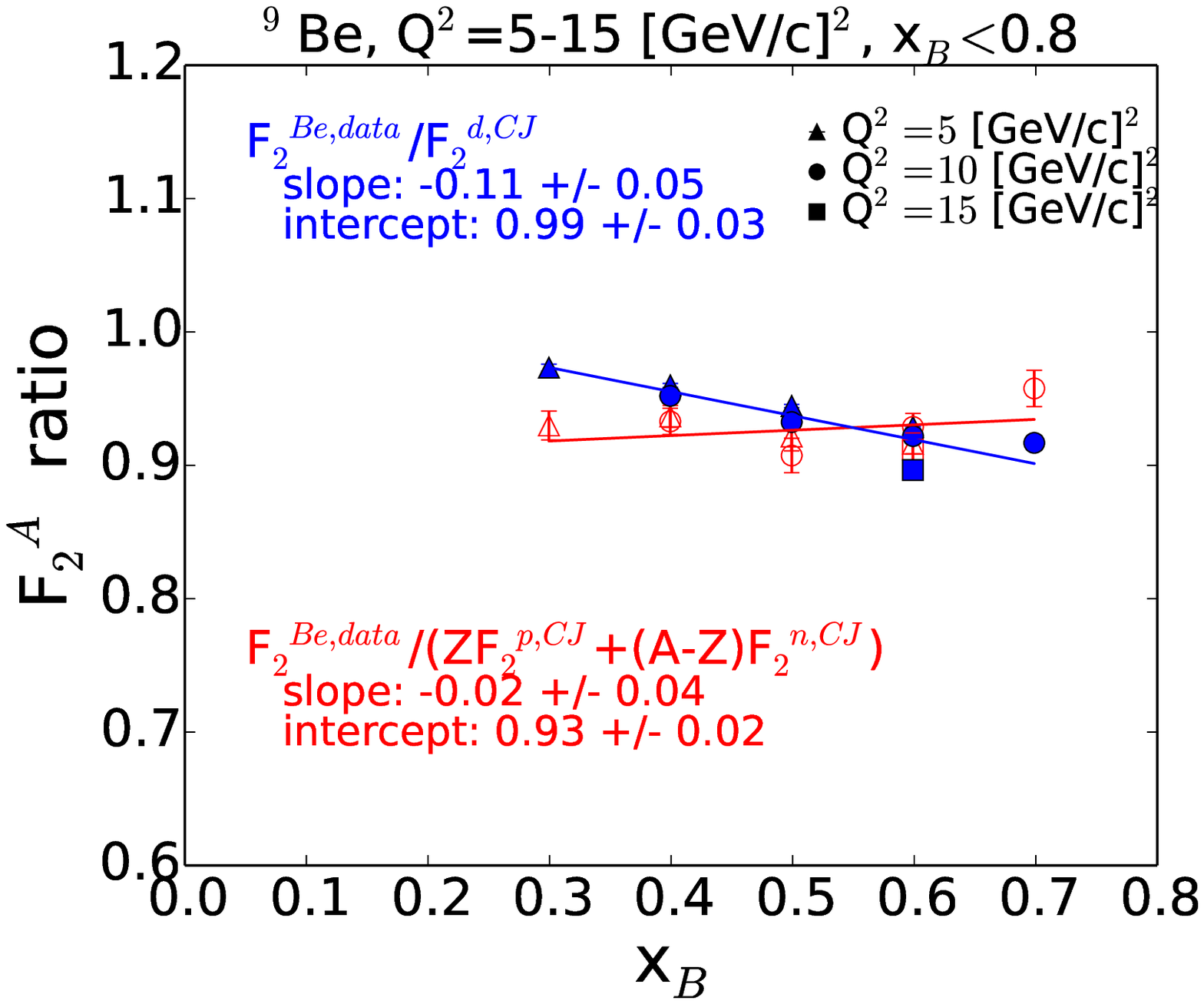}
\end{minipage}
  \caption[]{Linear fits to the $Be$ target data with cuts on $Q^2$ and $x_B$ from the E139 experiment. (a) The ratios are constructed requiring that $Q^2=5$~GeV$^2$/c$^2$. (b) The ratios include all data $5<Q^2<15$~GeV$^2$/c$^2$ where $x_B<0.7$. (c) The ratios include all data $5<Q^2<15$~GeV$^2$/c$^2$ up to $x_B<0.8$.}
  \label{fig:fits_Be}
\end{figure}

 \begin{figure}[H]
\begin{minipage}{0.46\textwidth}
\setlabel{pos=see,fontsize=\scriptsize,labelbox}
 \xincludegraphics[width=\textwidth,label=(a),labelbox=false,fontsize=\Large]{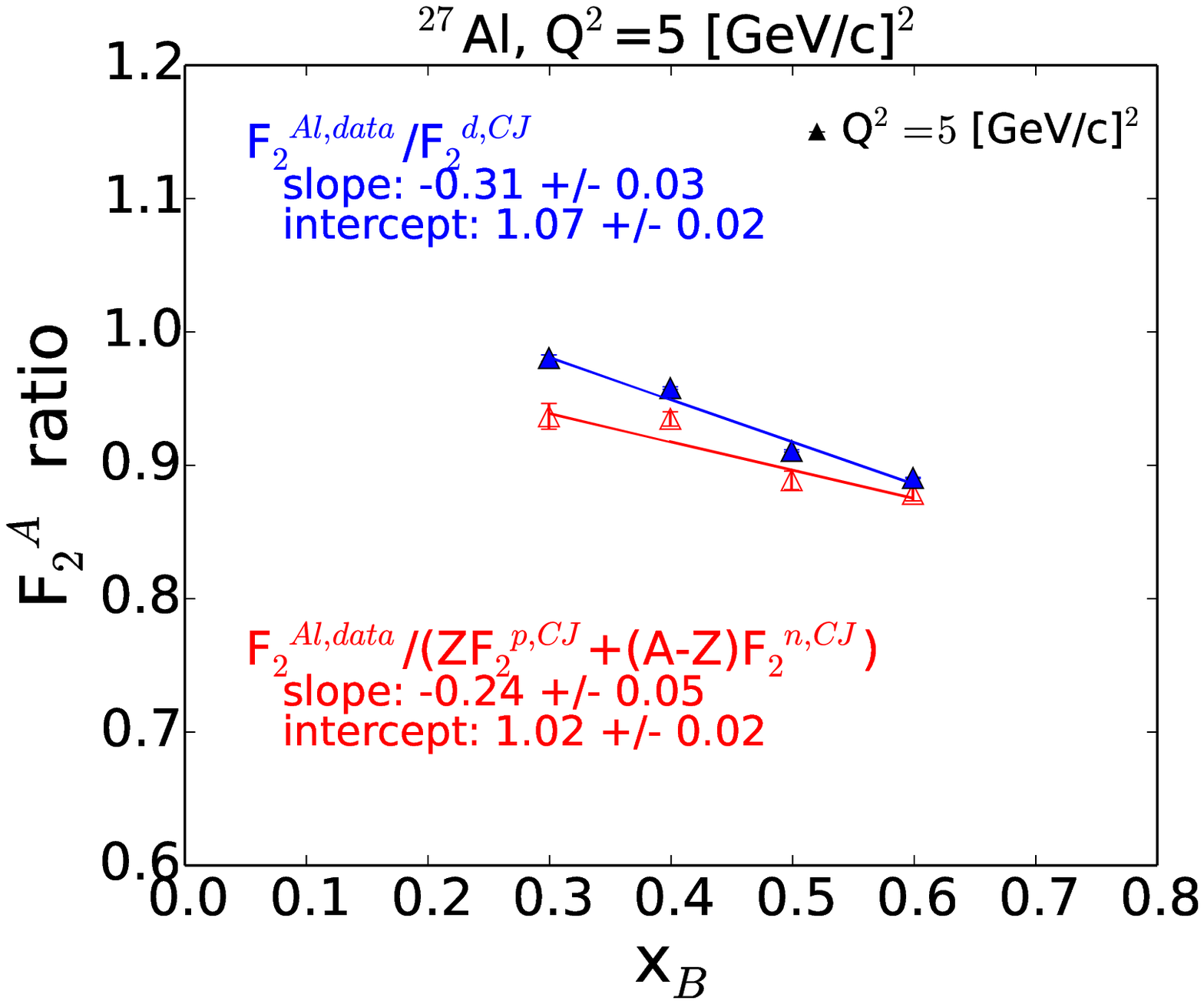}
\end{minipage}\hfill\begin{minipage}{0.46\textwidth}
\setlabel{pos=see,fontsize=\scriptsize,labelbox}
\xincludegraphics[width=\textwidth,label=(b),labelbox=false,fontsize=\Large]{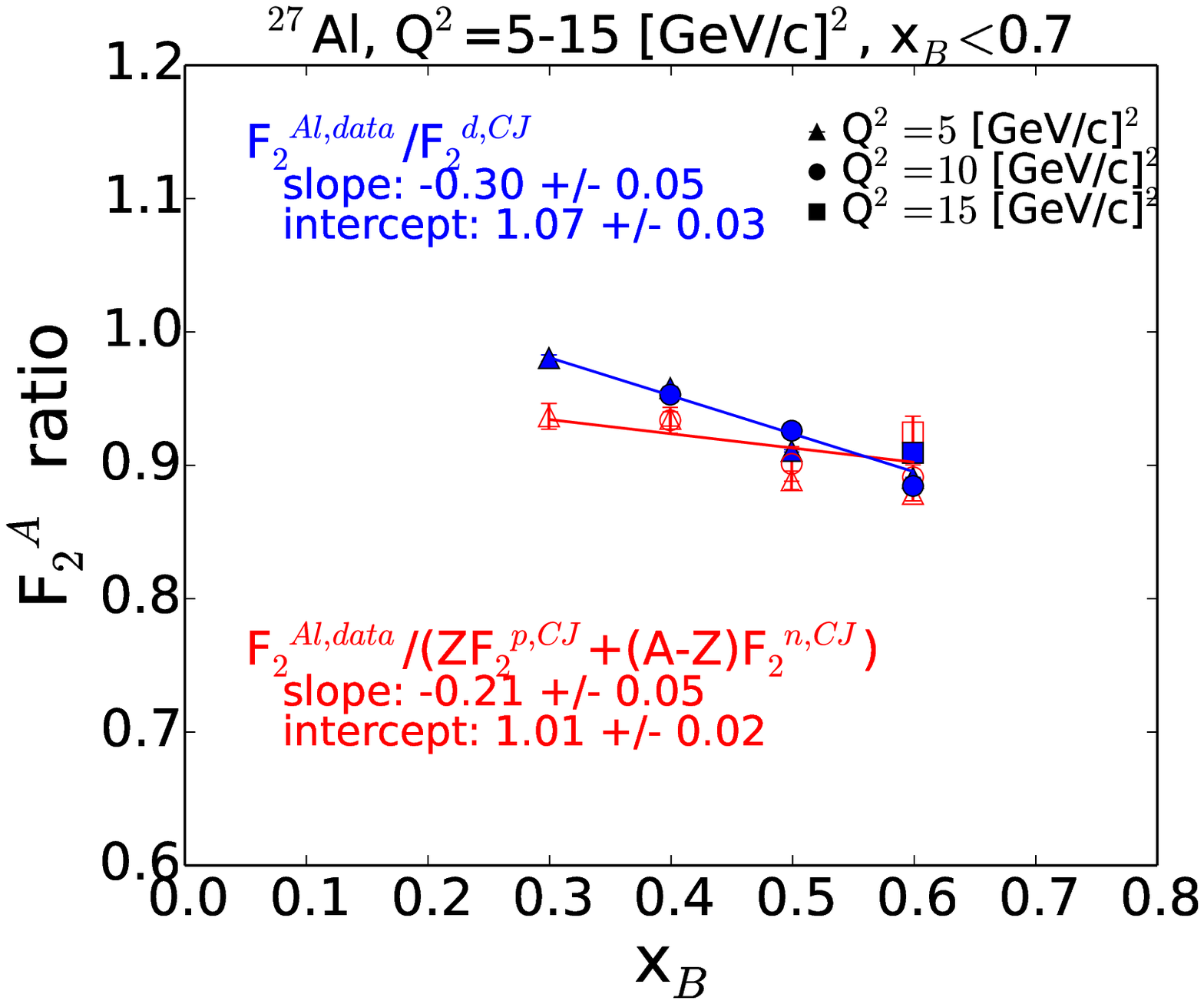}
\end{minipage}\hfill\begin{minipage}{0.46\textwidth}
\setlabel{pos=see,fontsize=\scriptsize,labelbox}
\xincludegraphics[width=\textwidth,label=(c),labelbox=false,fontsize=\Large]{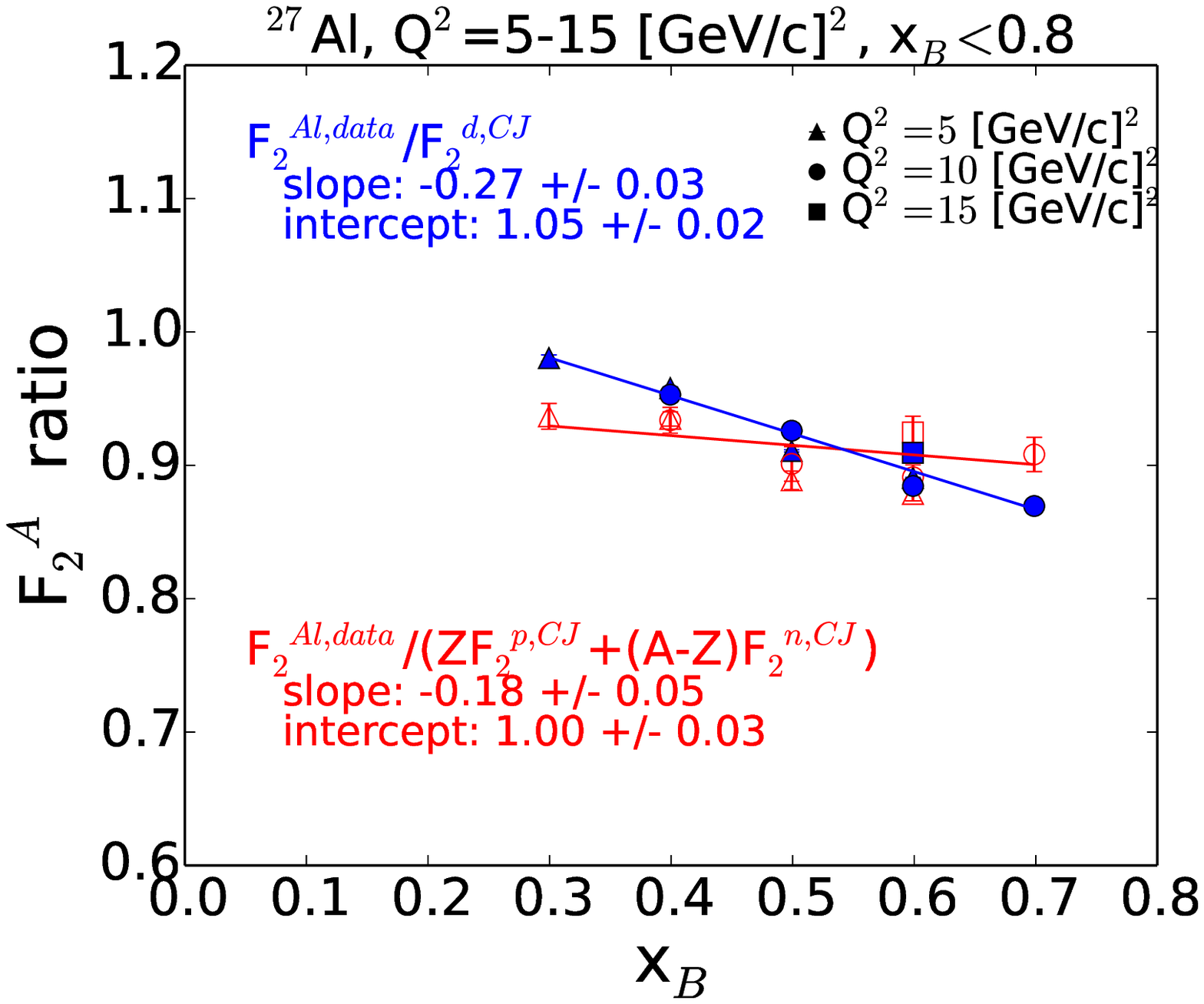}
\end{minipage}
  \caption[]{Linear fits to the $Al$ target data with cuts on $Q^2$ and $x_B$ from the E139 experiment. (a) The ratios are constructed requiring that $Q^2=5$~GeV$^2$/c$^2$. (b) The ratios include all data $5<Q^2<15$~GeV$^2$/c$^2$ where $x_B<0.7$. (c) The ratios include all data $5<Q^2<15$~GeV$^2$/c$^2$ up to $x_B<0.8$.}
  \label{fig:fits_Al}
\end{figure}

 \begin{figure}[H]
\begin{minipage}{0.46\textwidth}
\setlabel{pos=see,fontsize=\scriptsize,labelbox}
 \xincludegraphics[width=\textwidth,label=(a),labelbox=false,fontsize=\Large]{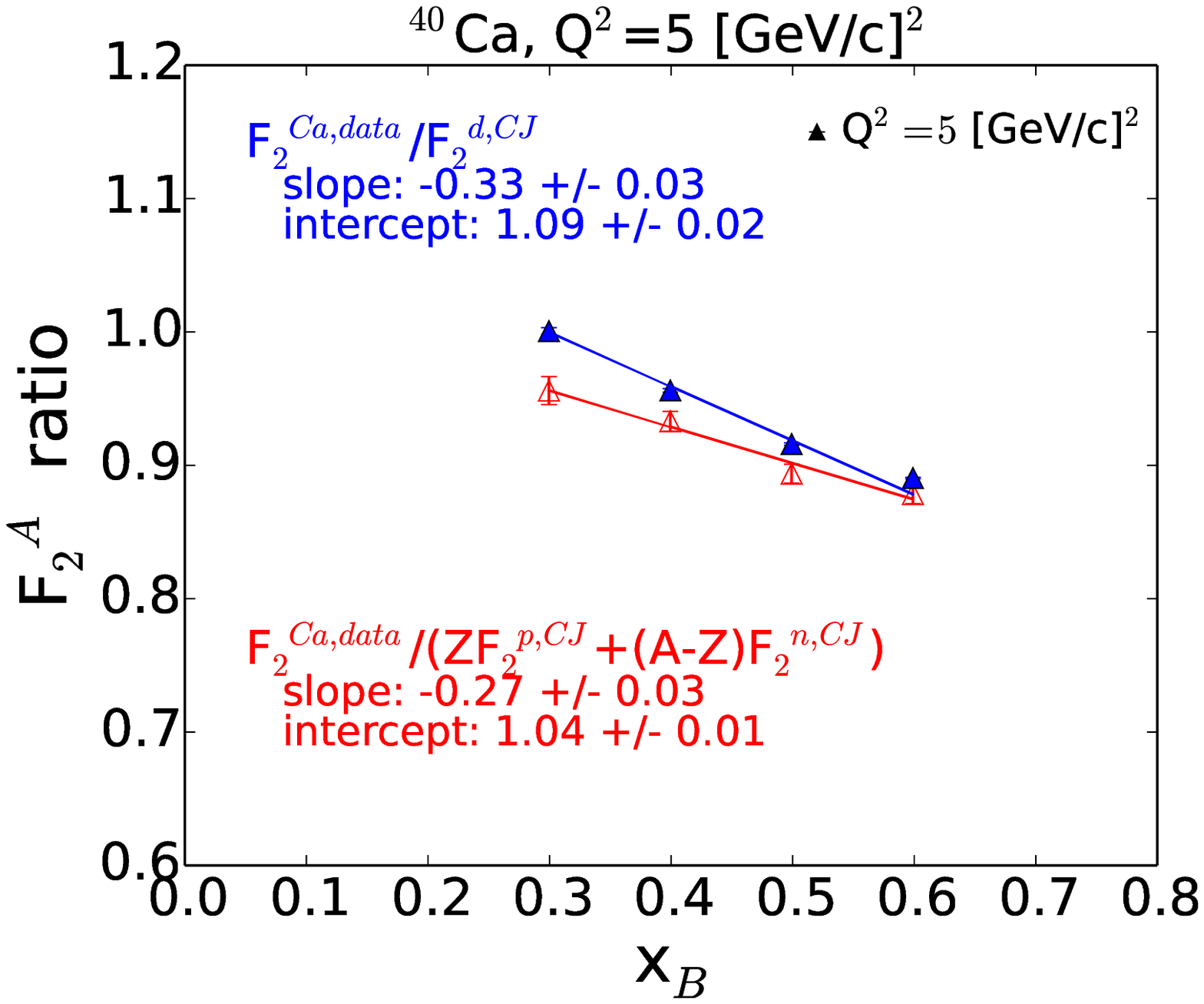}
\end{minipage}\hfill\begin{minipage}{0.46\textwidth}
\setlabel{pos=see,fontsize=\scriptsize,labelbox}
\xincludegraphics[width=\textwidth,label=(b),labelbox=false,fontsize=\Large]{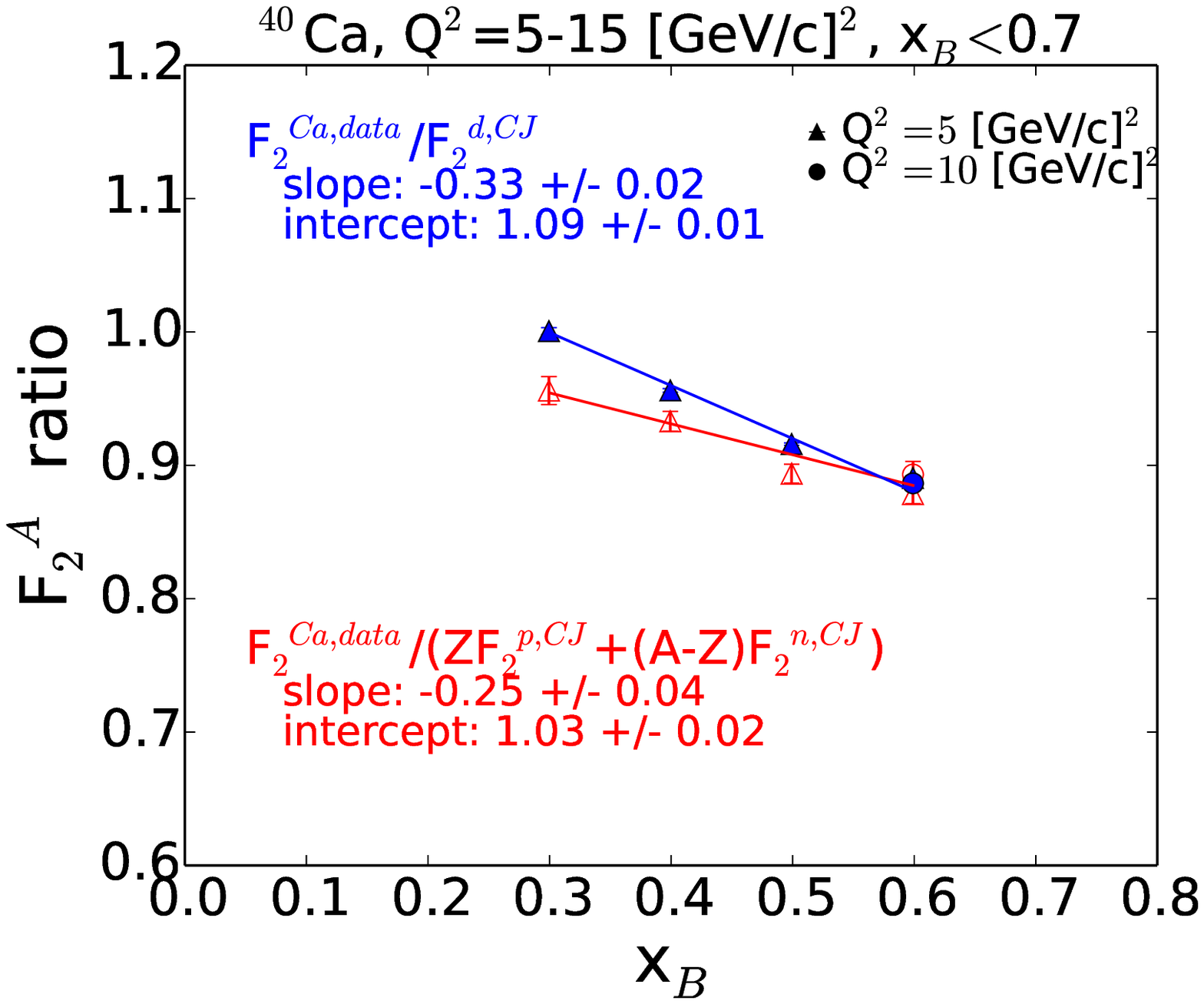}
\end{minipage}
  \caption[]{Linear fits to the $Ca$ target data with cuts on $Q^2$ and $x_B$ from the E139 experiment. (a) The ratios are constructed requiring that $Q^2=5$~GeV$^2$/c$^2$. (b) The ratios include all data $5<Q^2<15$~GeV$^2$/c$^2$ where $x_B<0.7$. No data was taken on $Ca$ beyond $x_B=0.6$.}
  \label{fig:fits_Ca}
\end{figure}   

 \begin{figure}[H]
\begin{minipage}{0.46\textwidth}
\setlabel{pos=see,fontsize=\scriptsize,labelbox}
 \xincludegraphics[width=\textwidth,label=(a),labelbox=false,fontsize=\Large]{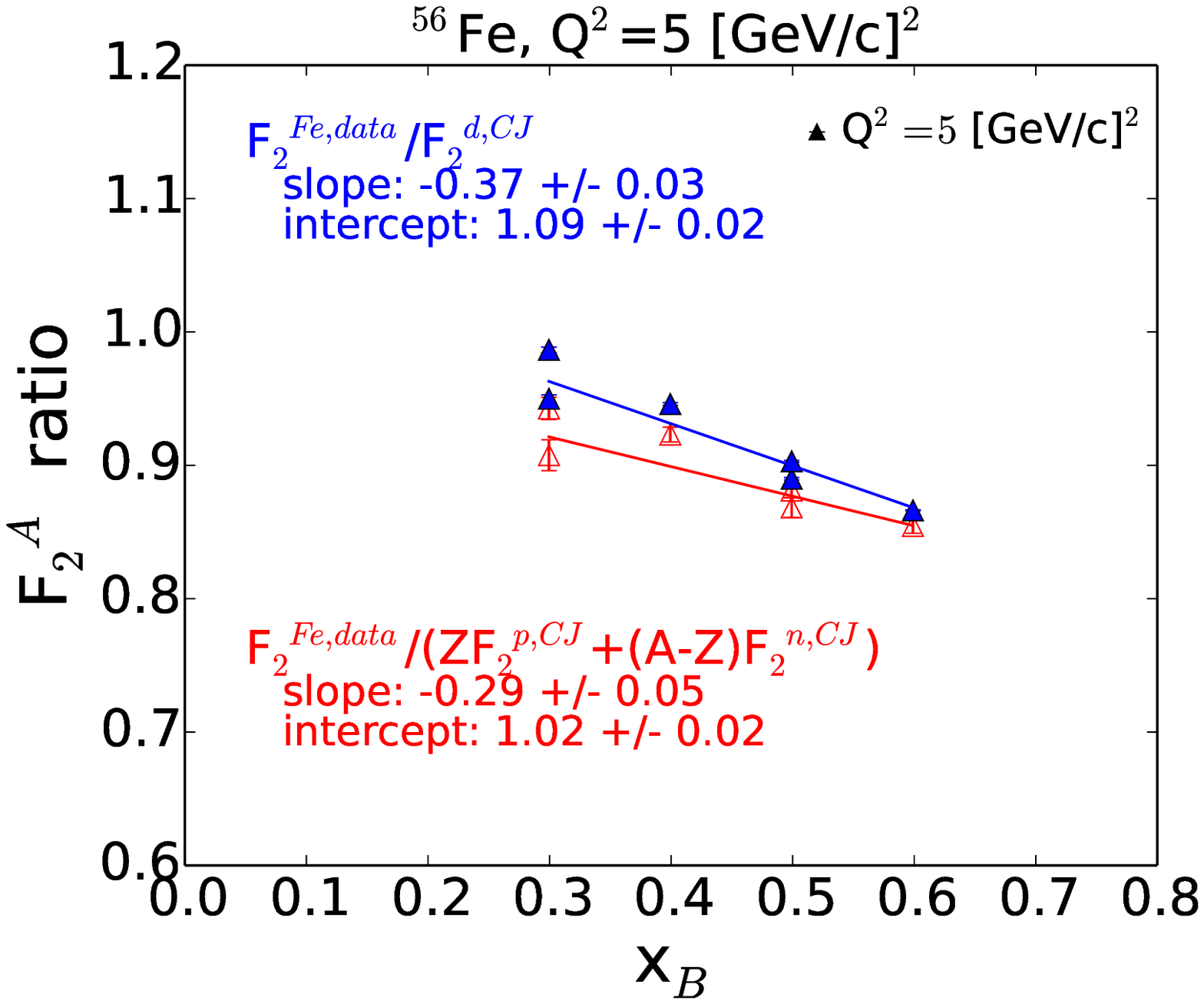}
\end{minipage}\hfill\begin{minipage}{0.46\textwidth}
\setlabel{pos=see,fontsize=\scriptsize,labelbox}
\xincludegraphics[width=\textwidth,label=(b),labelbox=false,fontsize=\Large]{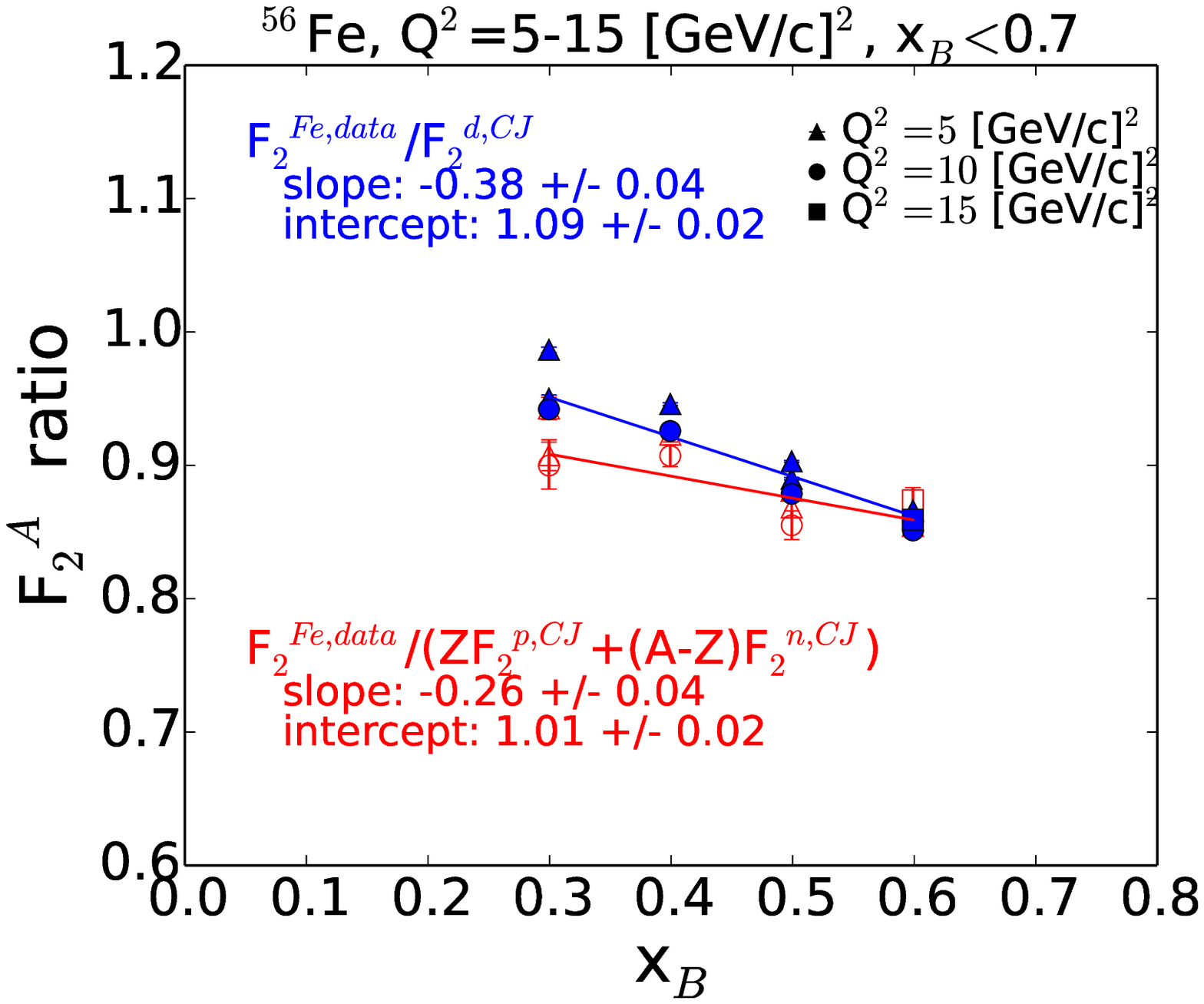}
\end{minipage}\hfill\begin{minipage}{0.46\textwidth}
\setlabel{pos=see,fontsize=\scriptsize,labelbox}
\xincludegraphics[width=\textwidth,label=(c),labelbox=false,fontsize=\Large]{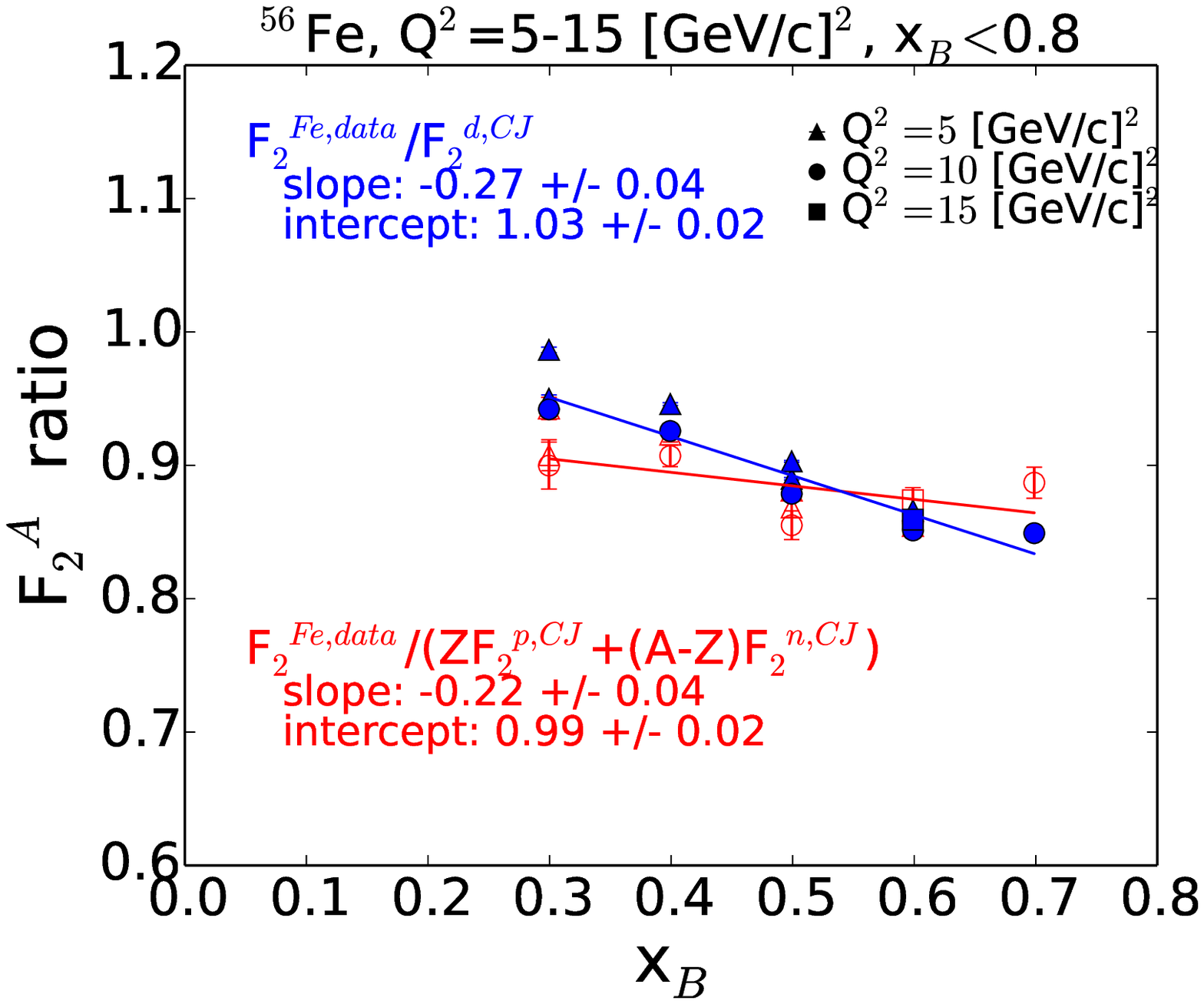}
\end{minipage}
  \caption[]{Linear fits to the $Fe$ target data with cuts on $Q^2$ and $x_B$ from the E139 experiment. (a) The ratios are constructed requiring that $Q^2=5$~GeV$^2$/c$^2$. (b) The ratios include all data $5<Q^2<15$~GeV$^2$/c$^2$ where $x_B<0.7$. (c) The ratios include all data $5<Q^2<15$~GeV$^2$/c$^2$ up to $x_B<0.8$.}
  \label{fig:fits_Fe}
\end{figure}   

 \begin{figure}[H]
 \centering
\begin{minipage}{0.46\textwidth}
\setlabel{pos=see,fontsize=\scriptsize,labelbox}
 \xincludegraphics[width=\textwidth,label=(a),labelbox=false,fontsize=\Large]{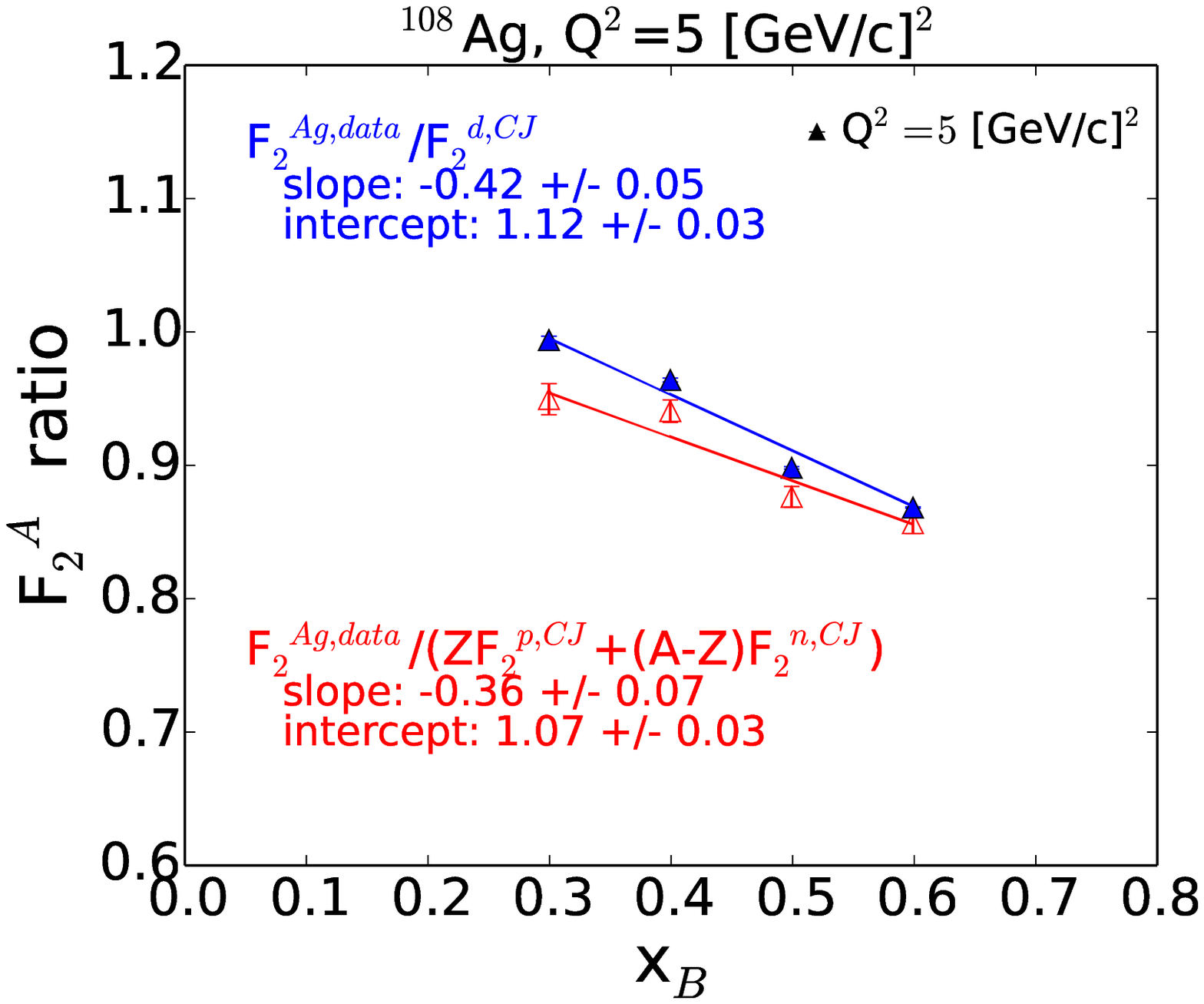}
\end{minipage}\hfill\begin{minipage}{0.46\textwidth}
\setlabel{pos=see,fontsize=\scriptsize,labelbox}
\xincludegraphics[width=\textwidth,label=(b),labelbox=false,fontsize=\Large]{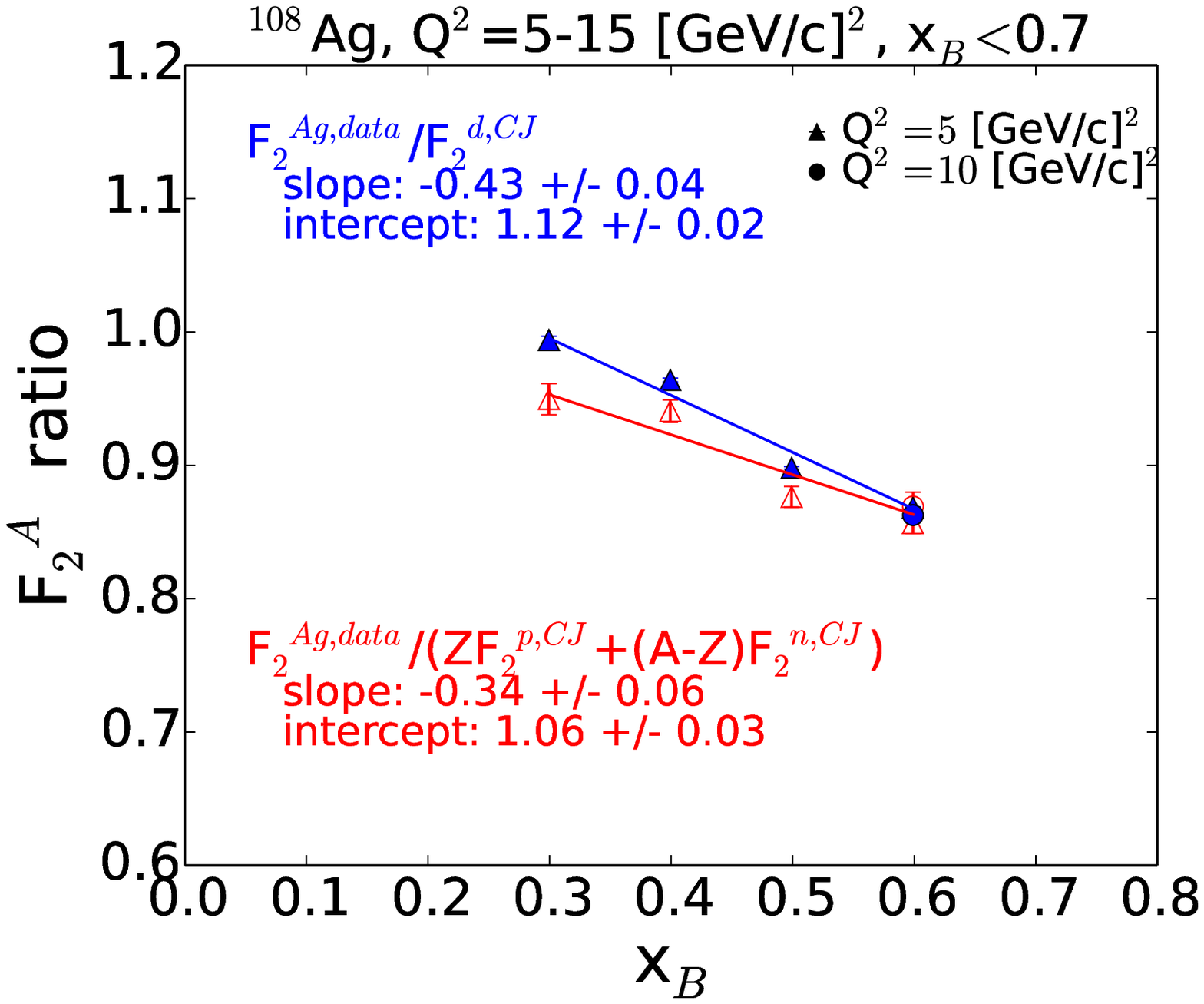}
\end{minipage}
  \caption[]{Linear fits to the $Ag$ target data with cuts on $Q^2$ and $x_B$ from the E139 experiment. (a) The ratios are constructed requiring that $Q^2=5$~GeV$^2$/c$^2$. (b) The ratios include all data $5<Q^2<15$~GeV$^2$/c$^2$ where $x_B<0.7$. No data was taken on $Ag$ beyond $x_B=0.6$.}
  \label{fig:fits_Ag}
\end{figure}   
 
\end{document}